\documentclass[preprint,journal]{vgtc}       

\usepackage{mathptmx}
\usepackage{mathrsfs}
\usepackage{graphicx}
\usepackage{times}
\usepackage{amsmath}
\usepackage{amssymb}
\usepackage{amsfonts}
\usepackage{algorithm}
\usepackage{subfigure}
\usepackage[american]{babel}
\usepackage{color, colortbl}
\usepackage{paralist}
\usepackage{booktabs}
\usepackage{multirow}
\usepackage{changepage}
\usepackage{moresize}
\usepackage[numbers,sort]{natbib}
\usepackage{algorithm}
\usepackage{algorithmicx}
\usepackage{balance}
\usepackage{algpseudocode}
\usepackage{soul}
\usepackage{xspace}
\usepackage{subfloat}
\usepackage{outlines}

\algrenewcommand\algorithmicrequire{\textbf{Input:}}
\algrenewcommand\algorithmicensure{\textbf{Output:}}

\definecolor{darkred}{RGB}{168,37,17}
\definecolor{darkblue}{RGB}{23,55,119}
\definecolor{highblue}{RGB}{20, 20, 180}
\definecolor{lightblue}{RGB}{183,210,237}
\definecolor{gray}{RGB}{100,100,100}
\definecolor{lightgray}{RGB}{230,230,230}
\definecolor{rainforest}{RGB}{3,101,100}
\definecolor{darkpurple}{RGB}{66,8,91}
\definecolor{orange}{RGB}{242, 101, 34}
\definecolor{black}{RGB}{0, 0, 0}

\usepackage{microtype}
\renewcommand{\paragraph}[1]{{\bf #1} }

\newcommand{\old}[1]{\xspace}

\usepackage[bookmarks,backref=true,linkcolor=black]{hyperref} 
\hypersetup{
  pdfauthor = {},
  pdftitle = {},
  pdfsubject = {},
  pdfkeywords = {},
  colorlinks=true,
  linkcolor= black,
  citecolor= black,
  pageanchor=true,
  urlcolor = black,
  plainpages = false,
  linktocpage
}

\onlineid{1160}

\vgtccategory{Research}
\vgtcpapertype{Algorithm/Technique}

\vgtcinsertpkg

\title{Retrieve-Then-Adapt: Example-based Automatic Generation for Proportion-related Infographics}

\author{Chunyao Qian, Shizhao Sun, Weiwei Cui, Jian-Guang Lou, Haidong Zhang, and Dongmei Zhang}
\authorfooter{
\item{S. Sun, W. Cui, J. Lou, H. Zhang, and D. Zhang are with Microsoft Research Asia. Emails: \{shizsu, weiweicu, jlou, haizhang, and dongmeiz\}@microsoft.com.}
\item{C. Qian is with Peking University, and this work was done during an internship at Microsoft Research Asia. Email: cyqvanilla@pku.edu.cn.}
}

\shortauthortitle{Biv \MakeLowercase{\textit{et al.}}: Global Illumination for Fun and Profit}

\abstract{
Infographic is a data visualization technique which combines graphic and textual descriptions in an aesthetic and effective manner. Creating infographics is a difficult and time-consuming process which often requires significant attempts and adjustments even for experienced designers, not to mention novice users with limited design expertise. Recently, a few approaches have been proposed to automate the creation process by applying predefined blueprints to user information. However, predefined blueprints are often hard to create, hence limited in volume and diversity. In contrast, good infogrpahics have been created by professionals and accumulated on the Internet rapidly. These \emph{online examples} often represent a wide variety of design styles, and serve as exemplars or inspiration to people who like to create their own infographics. Based on these observations, we propose to generate infographics by automatically imitating examples. We present a two-stage approach, namely retrieve-then-adapt.
In the \emph{retrieval} stage, we index online examples by their visual elements.
For a given user information, we transform it to a concrete query by sampling from a learned distribution about visual elements, and then find appropriate examples in our example library based on the similarity between example indexes and the query.
For a retrieved example, we generate an initial drafts by replacing its content with user information.
However, in many cases, user information cannot be perfectly fitted to retrieved examples.
Therefore, we further introduce an \emph{adaption} stage.
Specifically, we propose a MCMC-like approach and leverage recursive neural networks to help adjust the initial draft and improve its visual appearance iteratively, until a satisfactory result is obtained.
We implement our approach on widely-used proportion-related infographics, and demonstrate its effectiveness by sample results and expert reviews.
}


\CCScatlist{ 
 \CCScat{K.6.1}{Management of Computing and Information Systems}%
{Project and People Management}{Life Cycle};
 \CCScat{K.7.m}{The Computing Profession}{Miscellaneous}{Ethics}
}

 \teaser{
  \centering
  \includegraphics[width = \linewidth]{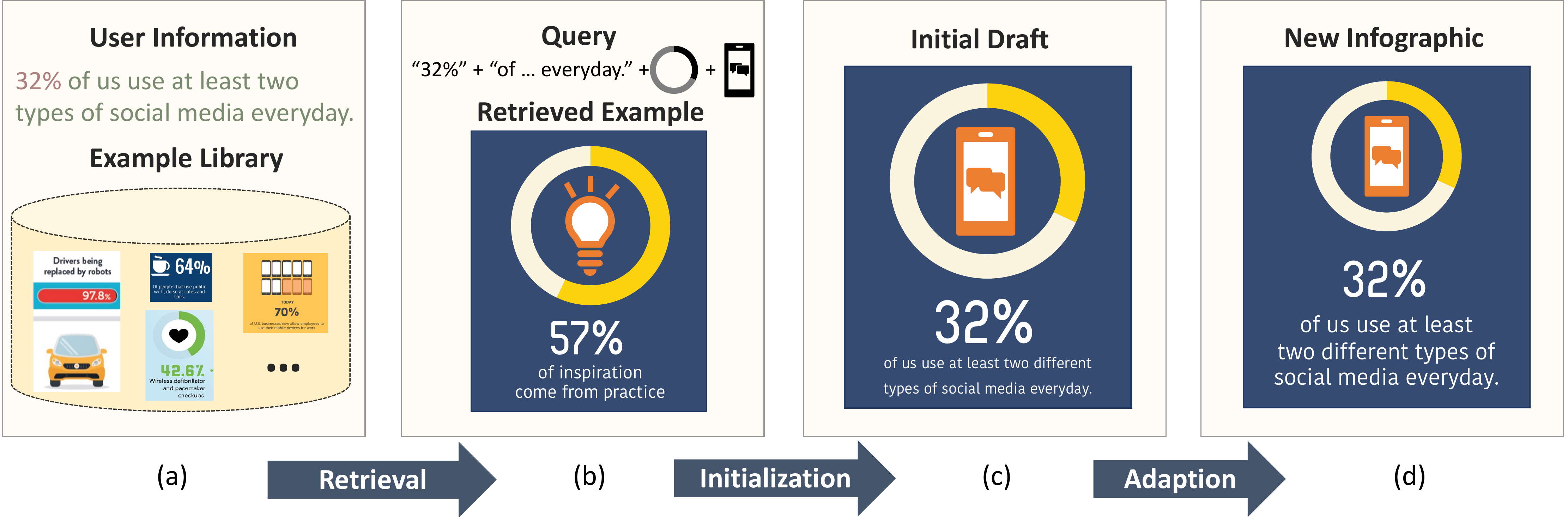}
 	\caption{An example-based approach for automatic infographics generation. a) The user information (top) and the example library (bottom) built by crawling infographics from the Internet. (b) The example (bottom) retrieved from the example library according to the query (top) generated from the user information. (c) The initial draft obtained by directly fitting user information into the design of the retrieved example. (d) The new infographic generated by adapting the design of the initial draft (e.g., enlarging the bottom text box).}
  \label{overview}
 }

\begin{document}
\maketitle

\section{Introduction}
\label{sec_intro}
Infographics~\cite{siricharoen2013infographics}, which often fuse text descriptions and graphic elements~\cite{purchase2018classification} to convey the information, are widely used in advertisements, posters, magazines, etc. Compared with plain texts, infographics are obviously easier to capture viewers' attention and help them quickly understand complex information~\cite{alrwele2017effects}. However, creating a successful infographic is not an easy task which requires professional skills and tremendous time. Novice users will easily get lost in the vast amount of design choices. Even experienced designers have to go through many iterations of adjustment before proposing a final infographic~\cite{herring2009getting}.

To lower the authoring barrier for casual users, predefined blueprints are widely used to automate the design of infographics in both commercial software (e.g., Microsoft PowerPoint and Adobe Illustrator) and new research paradigms (e.g., Text-to-Viz~\cite{cui2019text}). However, such blueprint-based approach has some limitations as well. First, as the number of predefined blueprints is limited, such approach may easily make the generated infographics limited and deja vu in terms of designs. Moreover, it is difficult to enrich the blueprint library with low efforts as designing blueprints is both complicated and laborious.
For example, to build a useful template, designers need to consider numerous factors: what visual elements are allowed in the template; how they are spatially arranged; what is the size of each element; what is the font or font size; what is the ideal length for a text element in it, etc.
These considerations are typically formulated as constraints in the template.
Strict constraints can ensure good results, but often reduce the diversity of the results generated by the corresponding template.
Loose constraints are more flexible, but can potentially yield strange or even wrong results.
Therefore, designers need to thoroughly experiment and ensure the template can deliver good results with any valid input.
On the other hand, there is a large collection of well-designed infographics on the Internet, called \emph{online examples} for ease of reference. Currently, they are widely used as exemplars or inspirations for people who would like to create their own infographics.
Compared to predefined blueprints, online examples are inherently rich in diversity and quantity.
More importantly, design choices in these examples are generally appropriate and viable in practice, as they are made and endorsed by professionals.


Based on the above observations, we seek to generate infographics by automatically imitating online examples. 
However, there are two critical challenges when developing such an example-based approach for infographics. First, in a large collection of examples, how to find appropriate ones whose designs can be transferred to a specific piece of information. Second, even we find such a matching online example, it is unlikely that the information is a perfect fit in terms of every design aspect. For instance, the text length in the user information may be different with that in the example. 
To address these challenges, we propose a two-stage approach, called \emph{retrieve-then-adapt}, to automatically convert a piece of information to its infographic equivalents.

In the \emph{retrieval} stage, we build a retrieval system that indexes examples with their visual elements, such as charts, icons, and texts.
Then, for a piece of information given by users, we need to effectively transform it to a concrete query to collect appropriate examples in the example library.
Specifically, we first deconstruct the information and identify candidate visual elements that are applicable to individual information components. 
Then, we encode the visual elements into a valid query to retrieve examples that are composed of similar elements.
Obviously, an information component may have multiple ways to present visually.
Therefore, to ensure an authentic diversity that is consistent with the common practice of designers, we generate viable queries based on the distribution about visual elements observed in the crowdsourced examples.
For example, if the proportion-related information is more frequently visualized using donut charts with icons than pure icons in the collected examples, our method will be more likely to issue a query about donuts with icons than pure icons, accordingly.


In the \emph{adaption} stage, we address the second challenge by automatically adjusting the designs in retrieved examples to make them more suitable for user information.
In most cases, the design of the retrieved example cannot be perfectly matched with user information in every aspect.
Fortunately, it provides a good start point to customize and refine to achieve an appealing result.
Specifically, we first directly adopt the design choices (e.g., position, color, font, etc.) from the example to create an initial draft for user information.
Then, we propose a MCMC-like approach~\cite{qi2018human} to iteratively make small changes to the design of the initial draft and gradually improve the visual quality over the iterations, until there is no better design can be proposed and a stable state is reached.
To help evaluate the improvement after each iteration and ensure visual quality improves monotonically, we leverage recursive neural networks~\cite{socher2011parsing,li2019grains} to encode hierarchical structures of infographics into hidden vectors and build a scorer upon them to compare the visual qualities before and after a small change is applied.

To evaluate our approach, we collect a real-world dataset about proportion-related infographics from the Internet and implement our approach based on them. We present sample results to qualitatively demonstrate the performance of our approach. Moreover, we interview four experts and collect their comments for our generated infographics. Both sample results and expert reviews demonstrate that our approach can generate diverse infographics by imitating online examples. 




\section{Related Work}

\subsection{Infographics}
Research work on infographics mainly includes three aspects, i.e., \emph{effect}, \emph{understanding} and \emph{auto-generation}. First, traditional studies focus on exploring what effects infographics have on humans. For example, Bateman et al.~\cite{bateman2010useful} compare embellished charts with plain ones by measuring interpretation accuracy and long-term recall. Haroz et al.~\cite{haroz2015isotype} find that using pictographs can improve the performance of information exchanging in terms of memorability, reading speed, and engagement. In addition, many studies are conducted to understand the underlying message or the visual structure of an infographic. For example, Bylinskii et al.~\cite{BylinskiiUnderstanding} adopt OCR techniques to assign hashtags to infographics for information retrieval purpose. Lu et al.~\cite{luexploring} find 12 different Visual Information Flow (VIF) patterns in infographics. 

Recently, there has been a growing interest in generating infographics. Many design tools~\cite{wang2018infonice,Satyanarayan2014Lyra,kim2016data} have been developed to facilitate the creation of infographics in an interactive manner. However, these tools cannot completely automate the creation process. Users are still required to understand advanced operations and concepts in these tools, and use their design expertise to make choices among numerous visual elements and attributes. To further automate the process, Cui et al.~\cite{cui2019text} propose generating infographics from natural language statements with predefined blueprints. However, predefined blueprints are expensive to create and easily homogenize designs. These limitations of predefined blueprints motivate us to consider leveraging online examples, which are easy to get and inherently diverse.

\subsection{Reusing Examples}
Examples play an important role in design practice. On the one hand, experienced designers usually get inspirations from examples because they can offer rich design styles~\cite{herring2009getting}: color schemes, visual expression, etc. On the other hand, for novice users, it is far easier to adjust an existing example than to create a design from scratch~\cite{Lee2010Designing}. Such critical role of examples inspires researchers to investigate reusing examples to create new designs in many different tasks, such as \emph{web pages design} and \emph{charts design}. In web page design, Bricolage~\cite{kumar2011bricolage} tries to match visually and semantically similar page elements between the content source and a web example, and then transfers the content of the source page to the best matching element in the example. Different from Bricolage that only considers one page, WebCrystal~\cite{Chang2012WebCrystal} extracts and combines styling information from different existing websites. In chart design, much attention has been paid to converting existing charts into reusable style templates that can be easily applied to new data sources. Harper et al.~\cite{harper2014deconstructing} present a pair of tools for deconstructing and restyling existing D3 visualizations. The deconstruction tool analyzes a D3 visualization by extracting the data, marks, and mappings between them. The restyling tool lets users modify the visual attributes of marks as well as the mappings from data to these attributes. The deconstruction approach is further extended to take into account non data-encoding marks and their visual attributes~\cite{harper2017converting,hoque2019searching}. 

All these existing approaches focus on how to reuse designs from existing examples. Our approach also falls into this category. However, we focus on a different type of data: infographics.
The unique design space of infographics proposes two critical challenges besides design extraction, i.e., how to find appropriate examples for a specific piece of information and how to adapt the example's design to make it more suitable for the information. 

\subsection{Automatic Layout Generation}
\label{subsec_layout_generation}
Our \textit{adaption} stage is inspired by existing automatic layout generation methods for traditional tasks, e.g., graphics design and indoor scene generation. Early approaches~\cite{o2014learning,o2015designscape,qi2018human} investigate Markov Chain Monte Carlo (MCMC) methods to generate layouts, which iteratively propose new layout and choose the better layout until no better layout can be proposed. The performance of those approaches is sensitive to layout evaluation metrics. However, these evaluation metrics are usually task-specific and heavily rely on human intuition, limiting them in modeling infographic design. Recently, deep neural networks have been explored for layout generations~\cite{zheng2019content,li2019layoutgan,wang2018deep,li2019grains,lee2019neural}. Those methods regard layout generation as a special case of image generation, and then leverage generative adversarial networks (GAN)~\cite{goodfellow2014generative} or variational auto-encoder (VAE)~\cite{doersch2016tutorial} to solve it. For example, LayoutGAN~\cite{li2019layoutgan} proposes a wireframe rendering layer to learn the alignment; READ~\cite{gadi2019read} leverages a recursive autoencoder to model highly structured layouts using less training data; ContentGAN~\cite{zheng2019content} synthesizes layouts by considering visual and textual semantics. Although those methods do not rely on handcrafted features, they often fail to achieve comparable performance as MCMC methods due to the limited training data.

The adaption stage in this work is similar to MCMC methods. However, instead of handcrafting evaluation metrics, we learn a recursive neural network from the example library about how to evaluate designs. 



\section{Preliminaries}
\label{sec_preliminary}

\begin{figure}[t]
	\centering
	\includegraphics[width=\columnwidth]{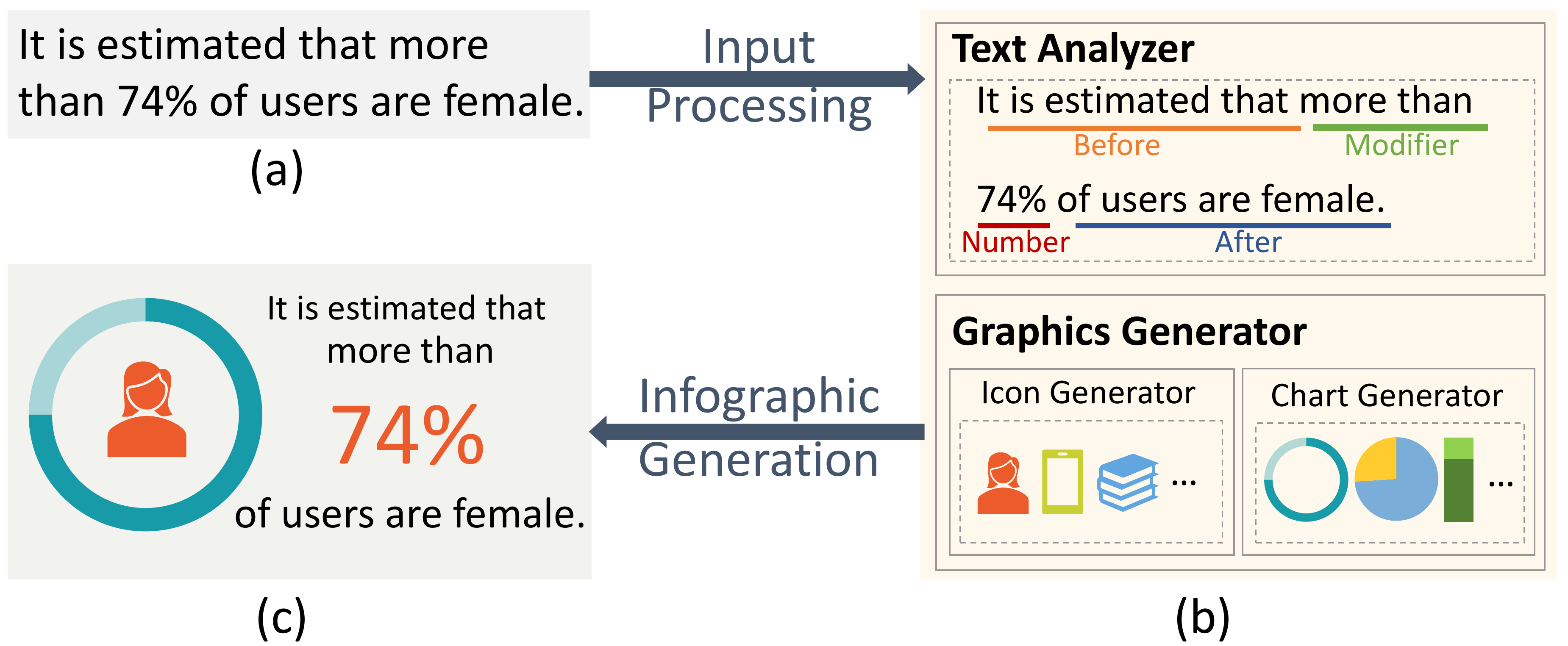}
	\caption{Example for our setting of infographics generation, where the input is a proportion-related natural language statement (see (a)) and the output is an infographic representing the information (see (c)). In this setting, (a) the input is first analyzed to obtain (b) various candidates of visual elements, which are then used to synthesize the infographic. 
	}
	\label{figs:task}
\end{figure}

Our approach is highly related to Text-to-Vis~\cite{cui2019text}.
In that work, Cui et al. first analyze and describe the text and visual spaces of proportion-related facts. 
Then, a blueprint-based solution is built to bridge these two spaces, which automatically converts a proportion-related natural language statement to its infographic equivalents.
According to their study, the motivation behind this input/output setting is two-fold.
First, the natural language is the most common way to convey information.
Second, the proportion-related infographic is the main category of infographics used in practice.

In this work, we adopt the same input/output setting of infographic generation (Figure~\ref{figs:task}), but on top of which we build our novel example-based approach.
In addition to the input/output setting, we also adopt the input processing pipeline (i.e., (a) to (b) in Figure~\ref{figs:task}) to help prepare queries for examples.
In this section, we briefly introduce these adopted algorithms and concepts that are used in the remainder of this paper.
More details can be found in the work of Text-to-Vis~\cite{cui2019text}.

%

\subsection{Input Processing}
\label{subsec_problem}
%
In this work, we follow the same input processing pipeline of Text-to-Vis to collect candidates of visual elements, since this module is not the core interest of this work and existing algorithms can already provide sufficient information to support our example-based approach.

Specifically, given a input statement that contains proportion-related information (e.g., Figure~\ref{figs:task}(a)), a \emph{text analyzer}, essentially a supervised CNN+CRF model, is utilized to split the statement into four different segments, namely, \emph{before}, \emph{modifier}, \emph{number}, and \emph{after}.
In addition, two separate \emph{graphic generators} are used to generate graphical elements. Specifically, an \emph{icon generator} is used to extract representative icons according to the semantics of the input statement from a predefined icon library and a \emph{chart generator} is used to generate pies, donuts, and bars according to the percentage value in the input statement.
All these text segments and graphics are candidate visual elements for information components in the input statement, and then will be used to construct viable queries for retrieving compatible examples in the example library (see Section~\ref{sec_retrieve}).

\subsection{Visual Elements}
\label{subsec_visual_space}
Following the visual space described in Text-to-Vis~\cite{cui2019text}, we rank and select the top ten frequently used visual elements in online examples.

\paragraph{Textual Elements.}
They are characterized by the contained information, which correspond with segments extracted by the \textit{text analyzer}.
\begin{compactitem}
	\item  \textbf{Before.} This refers to simple clauses in forms like ``It is estimated that'', ``Compared with last month'' or ``By the year 2020,''. They always appear at the beginning of the statement.
	\item  \textbf{Modifier.} This refers to adjunct words or phrases used before a number, such as ``only'', ``more than'', ``around'' or ``nearly''.
	\item  \textbf{Number.} This refers to the most important numerical information in a proportion-related fact. They have obvious textual patterns like ``10\%'', ``1 in 10'' ,``1 out of 10'' or ``1/10''.
	\item  \textbf{After.} This refers to the remaining texts after the number. For example, ``\dots of companies will be using artificial intelligence for driving digital revenue'' or ``\dots of people would like to receive promotions on social media''.
	\item  \textbf{Statement.} An input utterance may exist in the infographic as one text block sometimes. We consider the text block as a \emph{statement} element because it dose not split the input utterance into meaningful segments.
\end{compactitem}

\paragraph{Graphical Elements.} They are also critical parts in infographics, targeting attractive visual effects to emphasize the key points in the underlying information.

\begin{compactitem}
	\item  \textbf{Single icon.} Elements of this type are semantically related to the input utterance. Since icons are not explicitly provide in the input, we use the aforementioned \textit{icon generator} (Figure~\ref{figs:task}) to collect meaningful icons to represent the input.
	\item \textbf{Donut}/\textbf{Pie}/\textbf{Bar.} Elements of these types encode the numerical information of the input in a vivid way. In our implementation, they are programmatically generated according to the numerical information by \emph{chart generator} (Figure~\ref{figs:task}).	
	\item \textbf{Pictograph.} Compared with the previous four elements, pictographs not only encode numerical information, but also emphasize the subject of the proportion-related fact. In our implementation, they are created with \textit{icon generator} as well.
\end{compactitem}

Attributes of these textual elements and graphical elements are manually labeled in examples, and then used to index examples in the example library (see Section~\ref{sec_retrieve}).

\begin{figure}[t]
    \centering
    \includegraphics[width=\columnwidth]{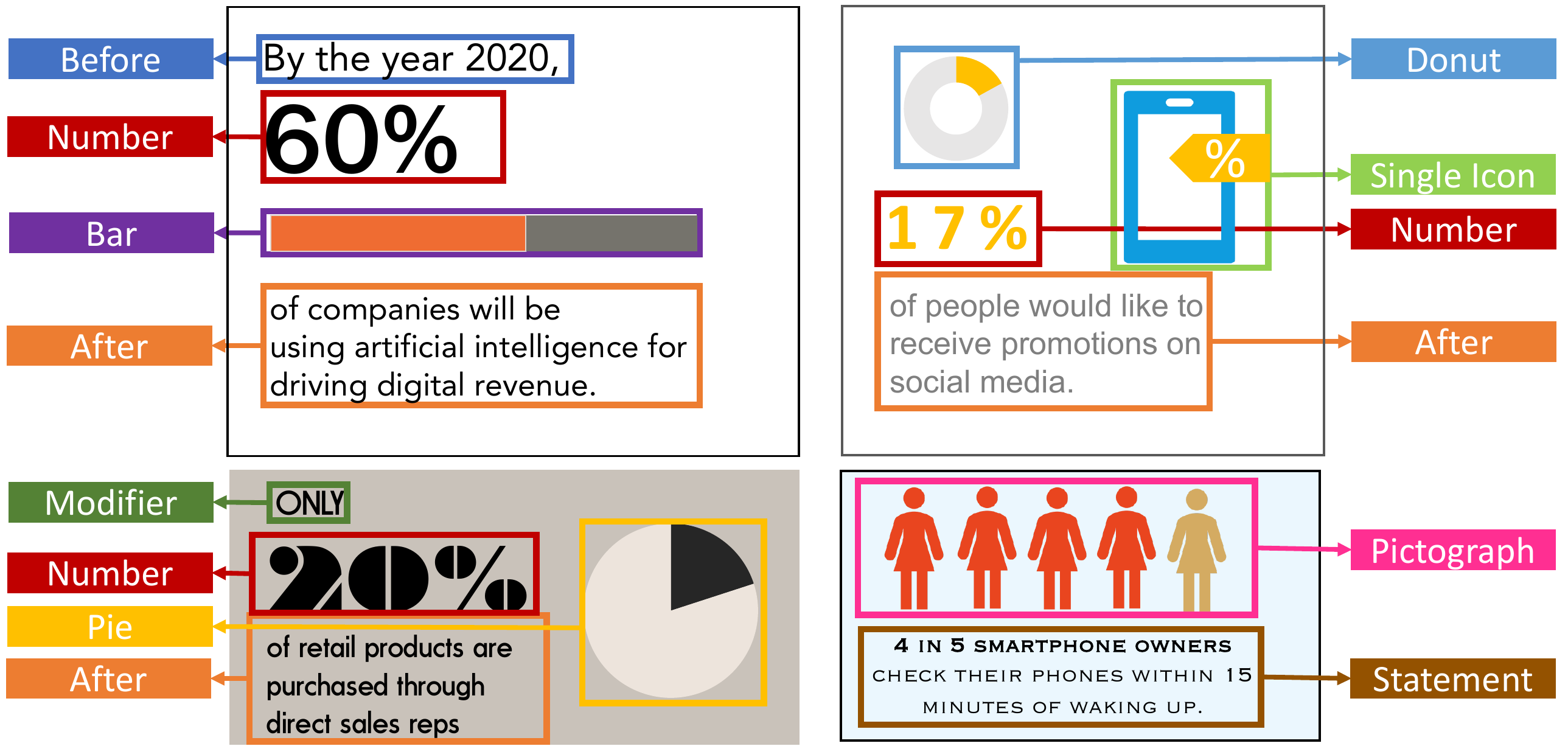}
    \caption{Examples for visual elements in proportion-related infographics.}
    \label{figs:visual_space}
\end{figure}

\section{System Overview}
\label{sec_overview}









Formally, we denote the example library as $\mathbb{D} = \{D_m\}_{m=1}^M$, which contains $M$ infographics $D_m$ crawled from the Internet. The input is denoted by $U$, which is a proportion-related natural language statement like ``More than 74\% of users are female''. Thus, our approach aims to automatically generate an infographic $G$ from $U$, by imitating designs in $\mathbb{D}$. Figure~\ref{overview} illustrates the workflow of our approach, which contains three key steps as follows.

\paragraph{Retrieval.} The first stage is to find appropriate examples whose designs can be transferred to the input $U$. We address this challenge by building a retrieval system that takes $U$ as the query to look up $\mathbb{D}$ and returns an appropriate example $\Tilde{D}\in\mathbb{D}$, i.e.,
$
    \Tilde{D} = f_{\text{retrieval}}\left(\mathbb{D}, U\right)
$.
Specifically, we index each example by its visual elements, e.g., charts, icons, and texts. Then, we transform an input $U$ to a concrete query in the same format of the example index. Finally, we retrieve examples based on the similarity between example indexes and queries (Figure~\ref{overview}(b)).

However, in many cases, $U$ has multiple valid infographic designs. 
For example, it can be visualized either as \textit{statement}+\textit{bar} or \textit{statement}+\textit{pictograph}, which are different in terms of queries.
To resolve this one-to-many mapping issue, we generate effective queries based on the distribution observed in the crowdsourced examples.
Intuitively, we assume the example library indicates the likelihood of different design choices that are adopted in practice. This motivates us to generate queries by first learning a distribution about visual elements from the example library and then sampling from the distribution. In this way, we ensure that examples are retrieved based on the popularity of their design choices in the real world. Moreover, by sampling from the learned distribution multiple times, it is possible to obtain different design choices (i.e., different concrete queries for the same $U$), leading to improved diversity of generated infographics. Figure~\ref{overview}(b) shows a concrete query which contains visual elements ``number+after+donut+single icon'' and the corresponding retrieved example.


\paragraph{Initialization.} We generate an initial draft $\Tilde{G}$ by directly applying the design of the retrieved example $\Tilde{D}$ to $U$, i.e.,
$
    \Tilde{G} = f_{\text{init}}\left(\Tilde{D}, U\right)
$.
We apply positions, colors, and text-specified attributes to the visual elements specified by the query (Figure~\ref{overview}(c)). If the retrieval stage returns multiple examples, we generate one initial draft for each example.

\paragraph{Adaption.} In many cases, directly applying the design may not work perfectly due to the mismatches between visual elements in the retrieved example and those used in the query. For example, text lengths or aspect ratios of icons may be slightly different, which may lead to a strange look in $\tilde{G}$, such as overlapping, excessive white space or improper font size (Figure~\ref{overview}(c)). We propose an \textit{adaption} stage to refine $\tilde{G}$ and make it more natural and suitable for the input $U$, i.e.,
$
   G = f_{\text{adaption}}\left(\Tilde{G}\right)
$.
Specifically, we propose a MCMC-like approach to adjust the design of the initial draft $\tilde{G}$ iteratively. First, a candidate design is proposed by randomly making a small change to the position or size of a visual element in the current design. Then, we compare the visual qualities of the current and candidate designs and choose the better one as the starting point for the next round of iteration.
However, evaluating designs is a nontrivial task as the criteria for a good design is intricate and hard to be quantified~\cite{harrington2004aesthetic,o2014learning}. To address this problem, we train a recursive neural network from the example library to evaluate the change in the aforementioned iteration. 


\section{Example Library Construction}
\label{subsec_datasets}

\begin{figure}[t]
    \centering
    \includegraphics[width=\columnwidth]{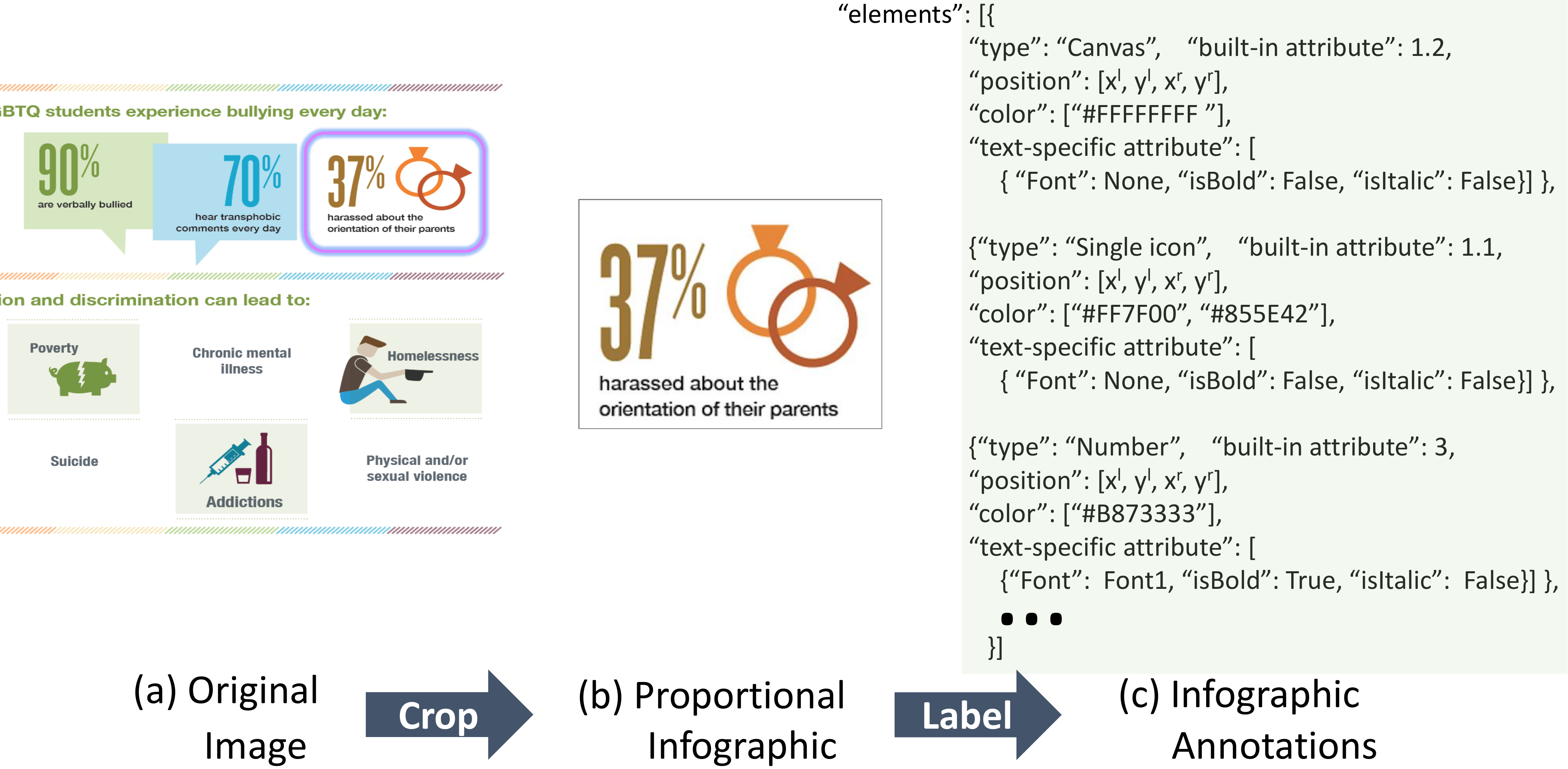}
    \caption{Illustration for the pipeline of example library construction.}
    \label{figs:dataset}
\end{figure}

The foundation of our work is a collection of online examples, which serves three purposes in our approach.
First, we analyze the examples to extract the distribution about visual elements.
Since there are no design preferences indicated in the input statement, we use the distribution to guide the query generation, so that the generated results have an authentic diversity consistent with real world examples.
Second, we index these examples in a database for our system to query and construct the initial draft for a given input. 
Third, we use the example corpus to train a recursive neural network that is used in the \textit{adaption} stage to evaluate the change at every iteration.

In this work, we build an example library by crawling and processing infographic exemplars from the Internet. First, we \emph{collect} a set of infographic sheets from the Internet. Then, as shown in Figure~\ref{figs:dataset}(a), an infographic sheet usually contains multiple infographic units. Therefore, we \emph{crop} the proportion-related units from each sheet. Finally, we \emph{label} the visual elements and their attributes in the cropped infographics. We introduce each step in detail as follows (Figure~\ref{figs:dataset}).

\paragraph{Collection.} 
We search Google Image by using a primary keyword ``infographic'', as well as a secondary keyword indicating a topic to ensure the diversity of dataset. Specifically, we consider ten common topics, i.e., education, health, commerce, ecology, diet, sports, animals, wedding, technology, and medical. In total, we downloaded 1000 infographic sheets with 100 under each topic.

\paragraph{Cropping.} 
Proportion-related infographics convey statistical information about how much a part occupies the whole, e.g., ``More than 74\% of users are female'', which provides obvious indications for us to crop them from the downloaded sheets.
Specifically, three coauthors review all the original sheets and crop proportion-related infographics from them independently. In total, we obtain 829 examples. 

\begin{figure}
\centering
\subfigure[]{
\includegraphics[width=0.71\linewidth]{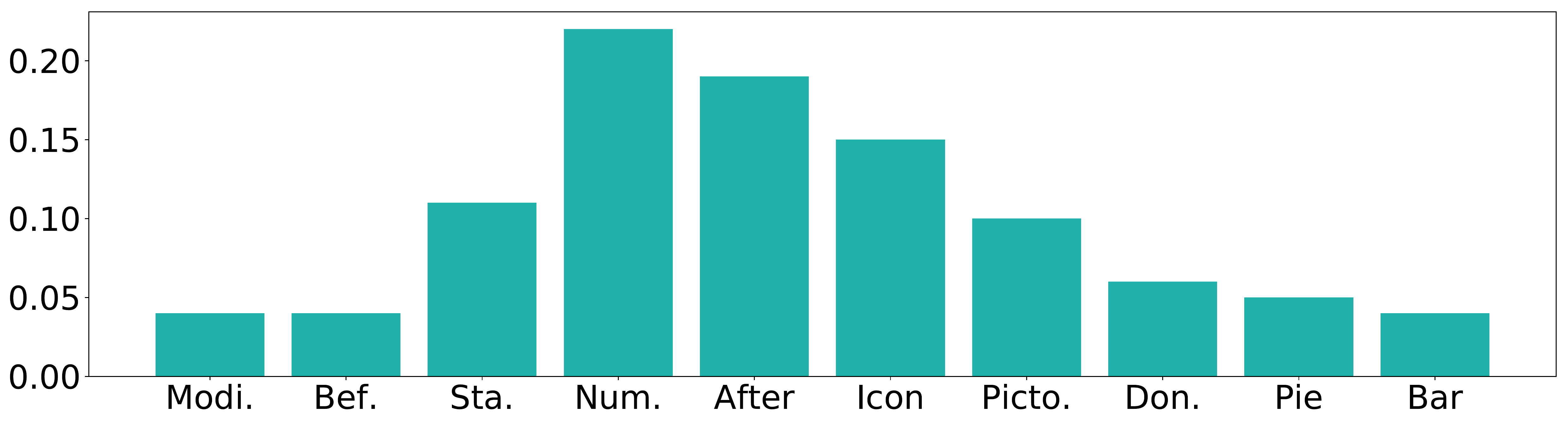}
}
\subfigure[]{
\includegraphics[width=0.24\linewidth]{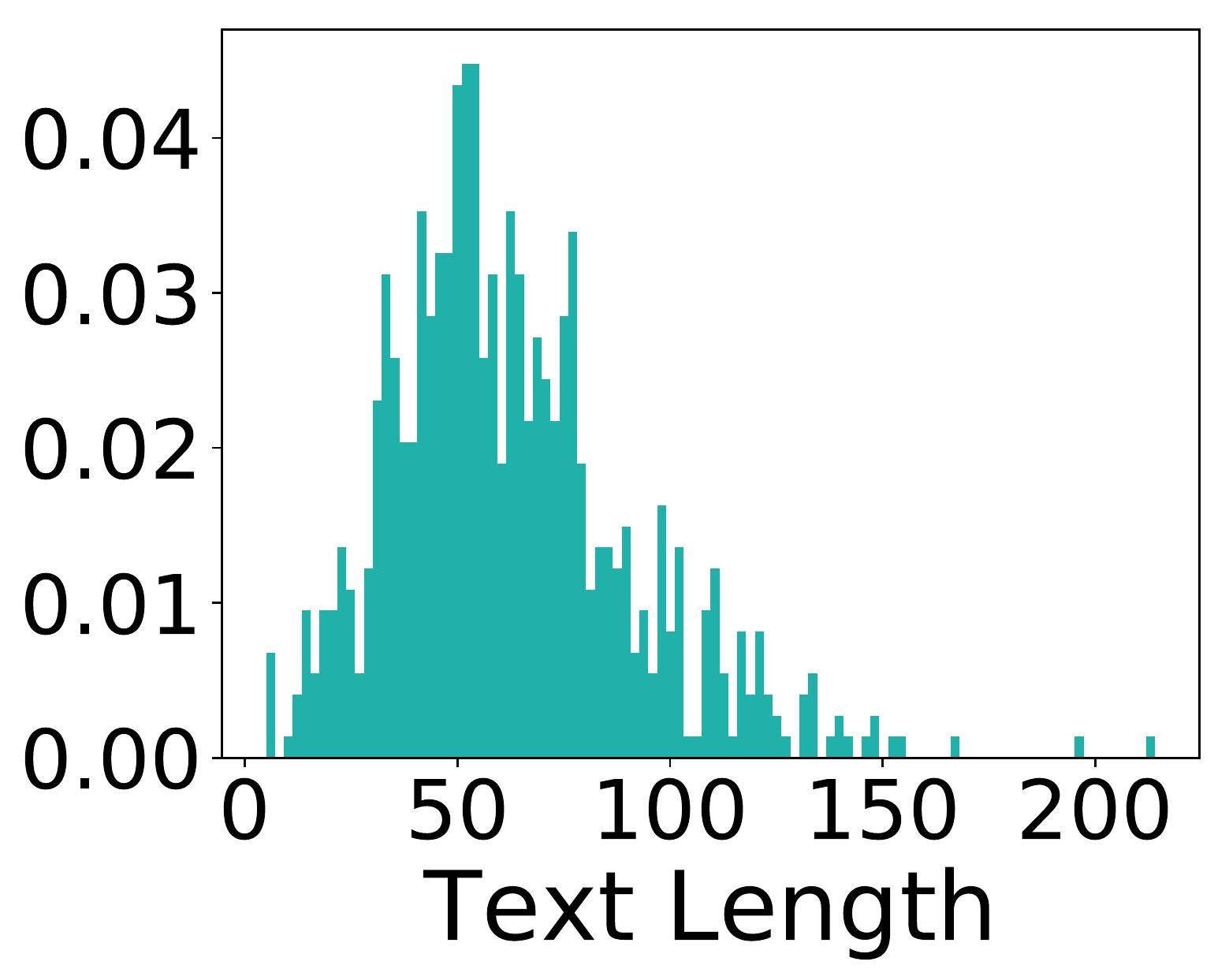}
}

\vspace{-1em}
\subfigure[]{
\includegraphics[width=0.17\linewidth]{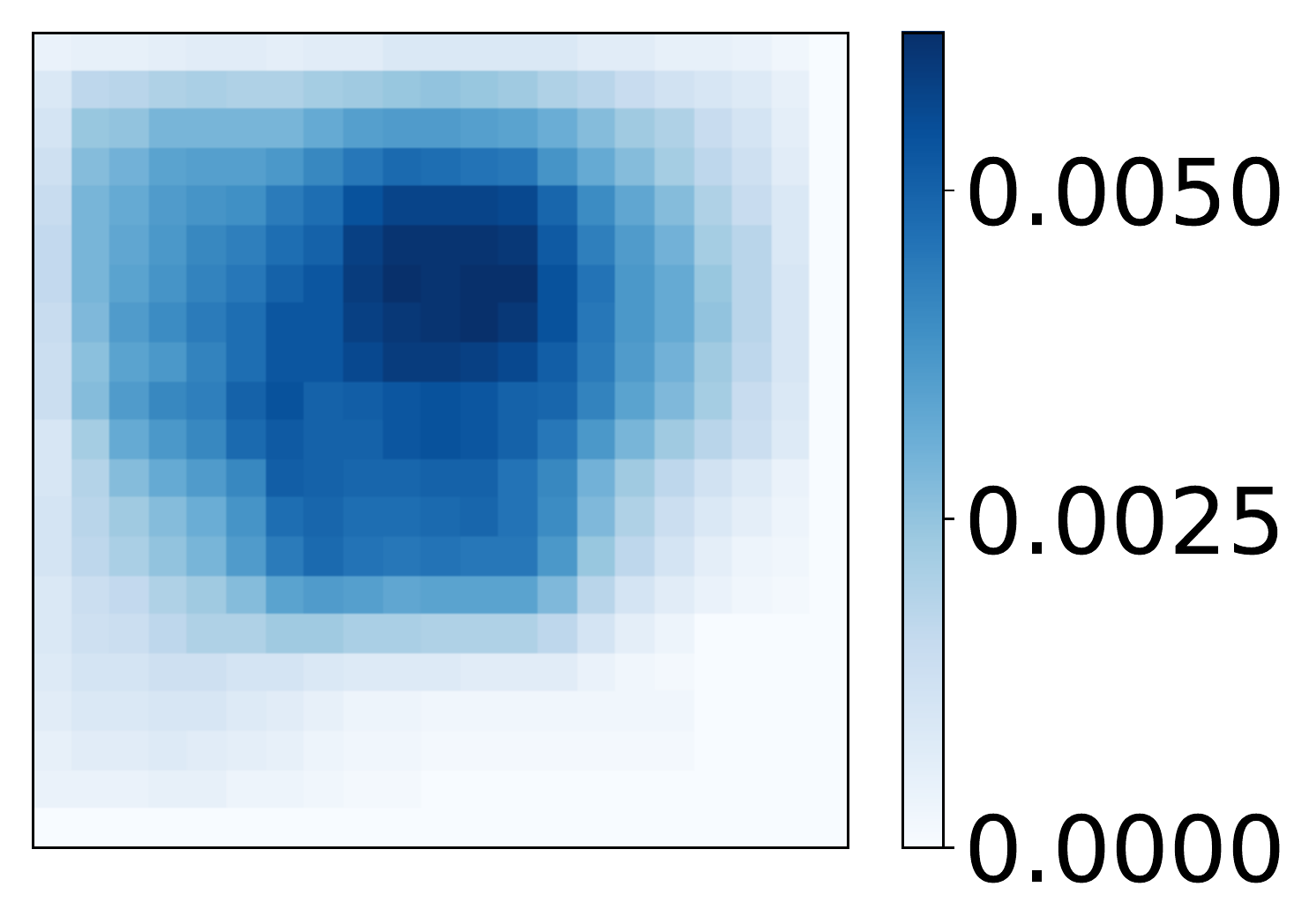}
}
\subfigure[]{
\includegraphics[width=0.17\linewidth]{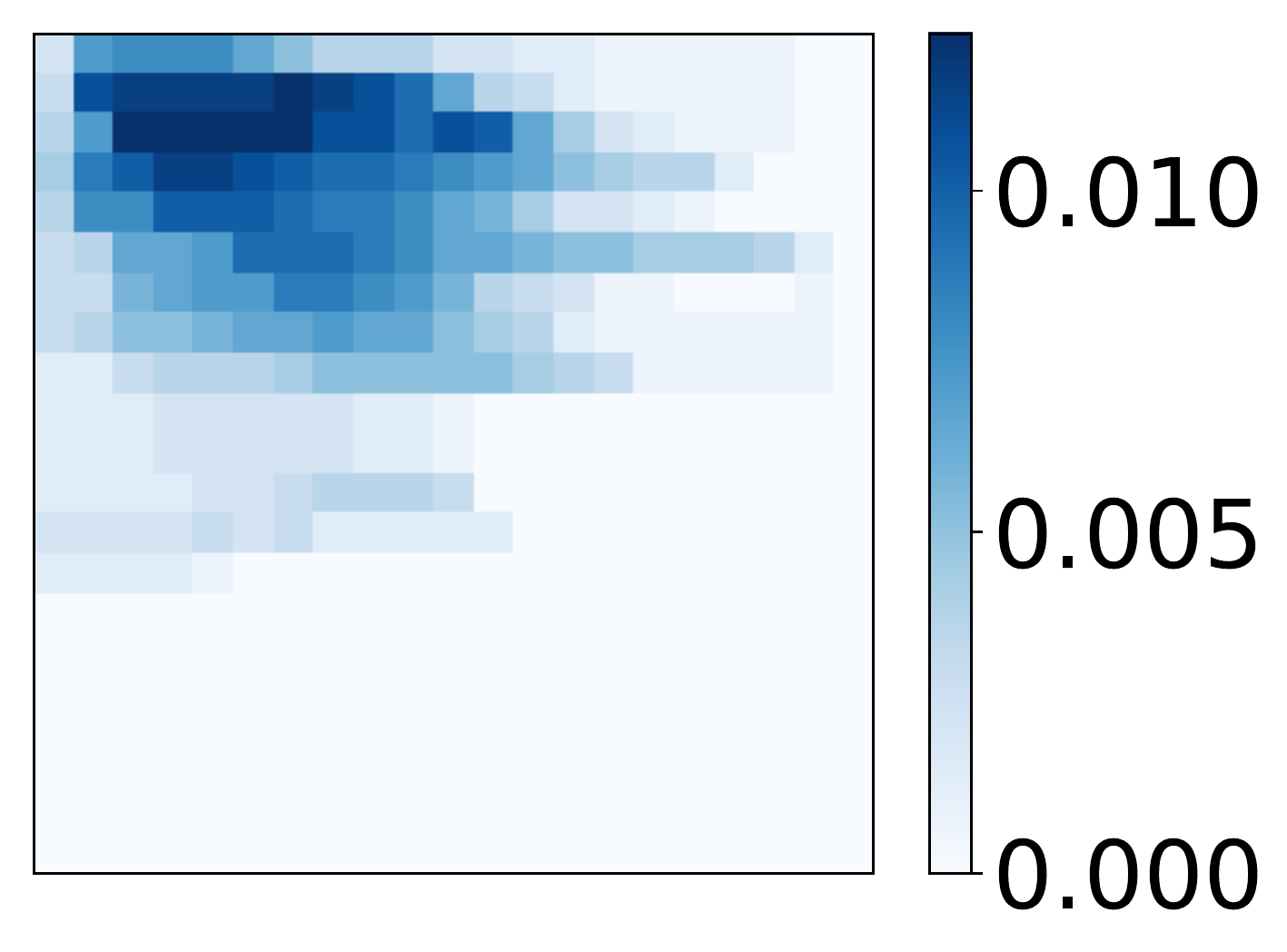}
}
\subfigure[]{
\includegraphics[width=0.17\linewidth]{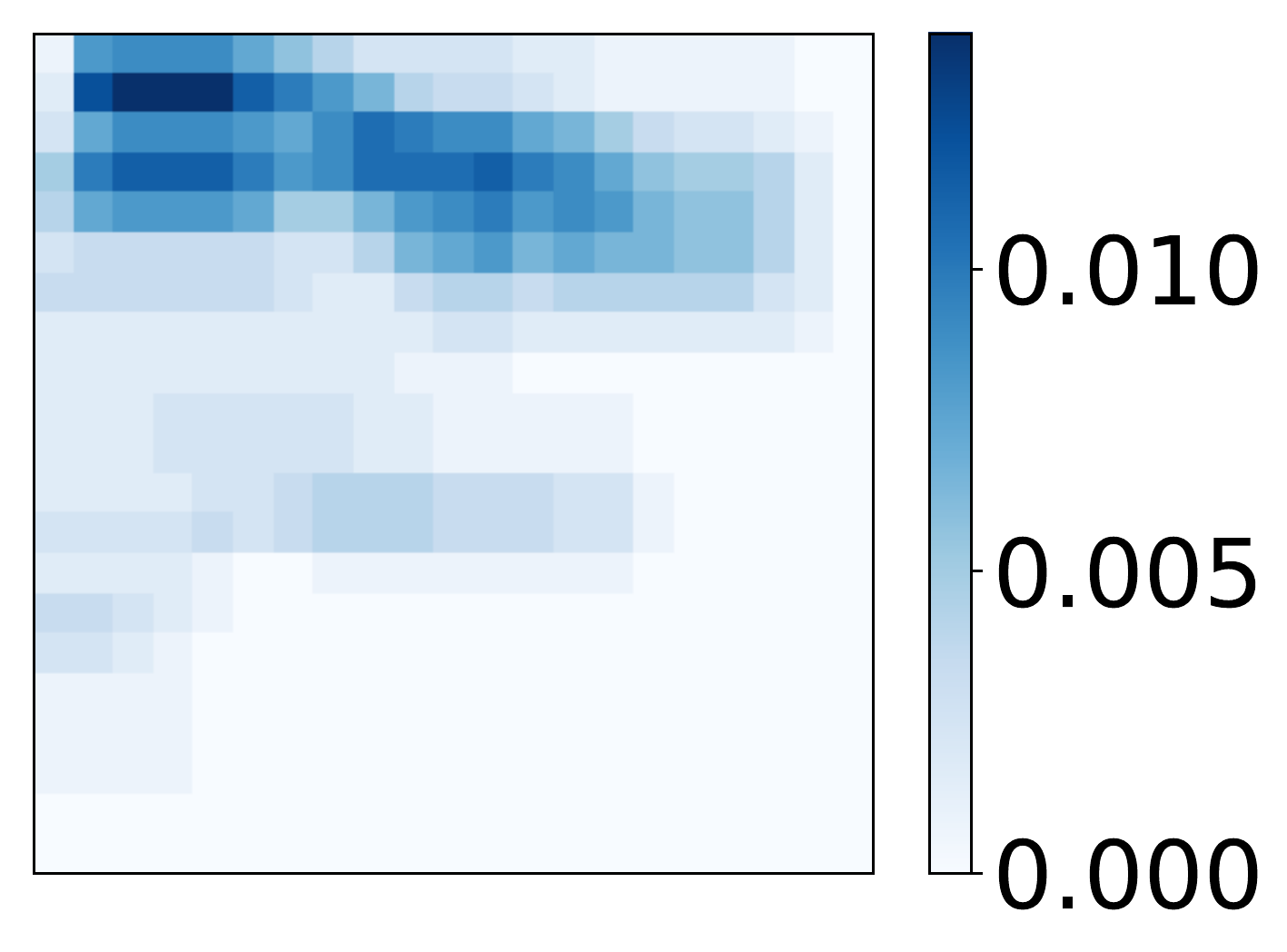}
}
\subfigure[]{
\includegraphics[width=0.17\linewidth]{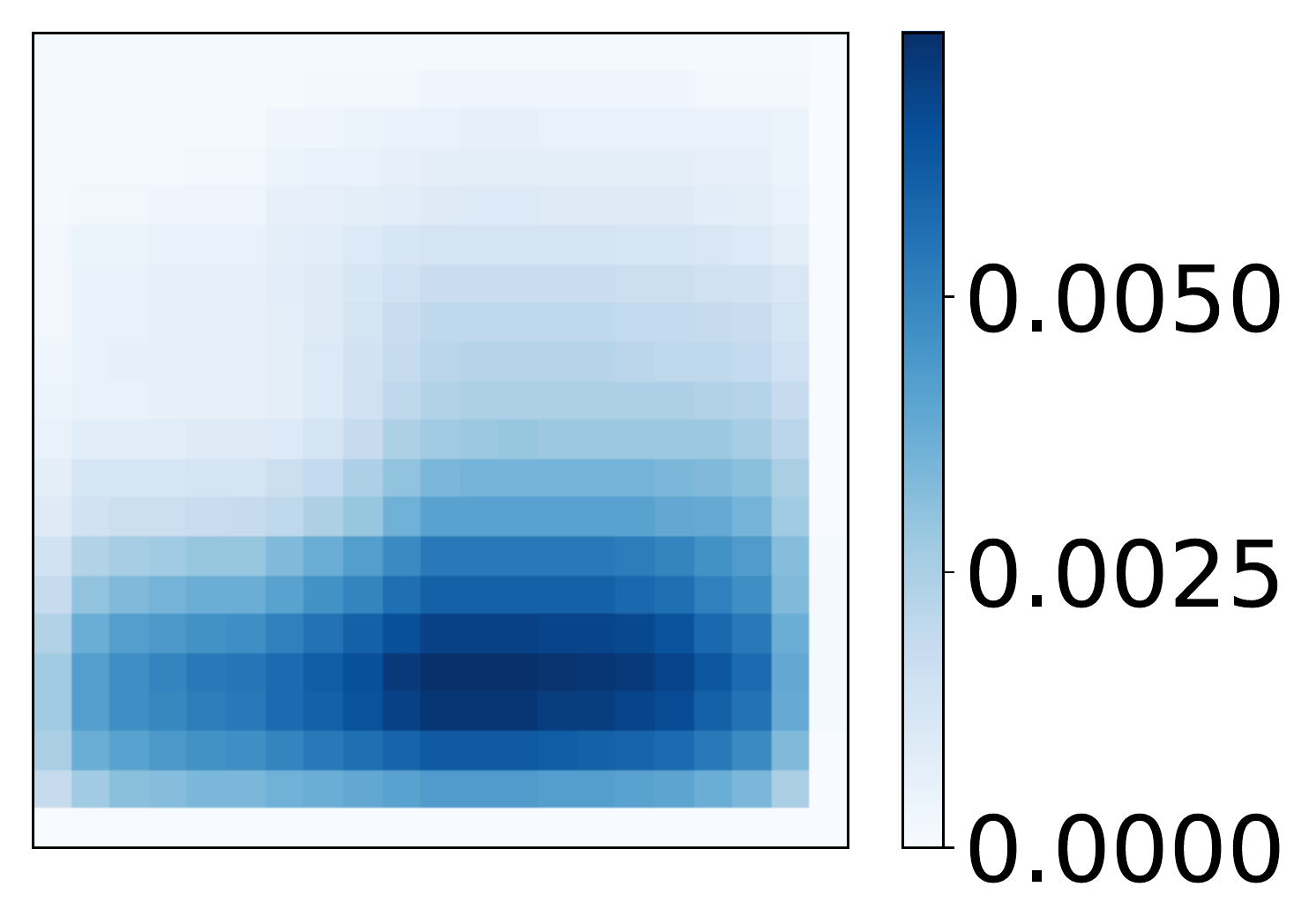}
}
\subfigure[]{
\includegraphics[width=0.17\linewidth]{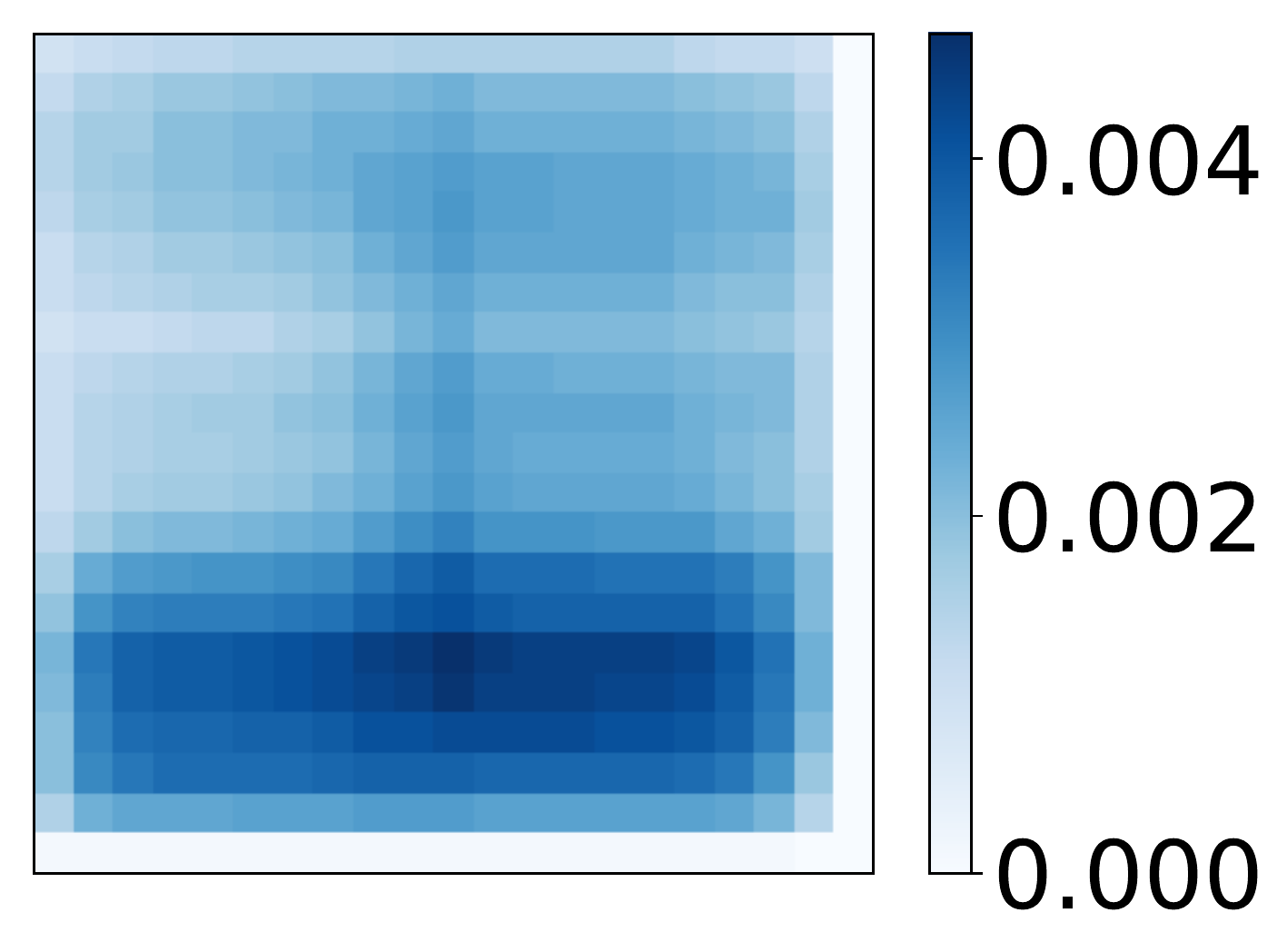}
}

\vspace{-0.9em}
\subfigure[]{
\includegraphics[width=0.17\linewidth]{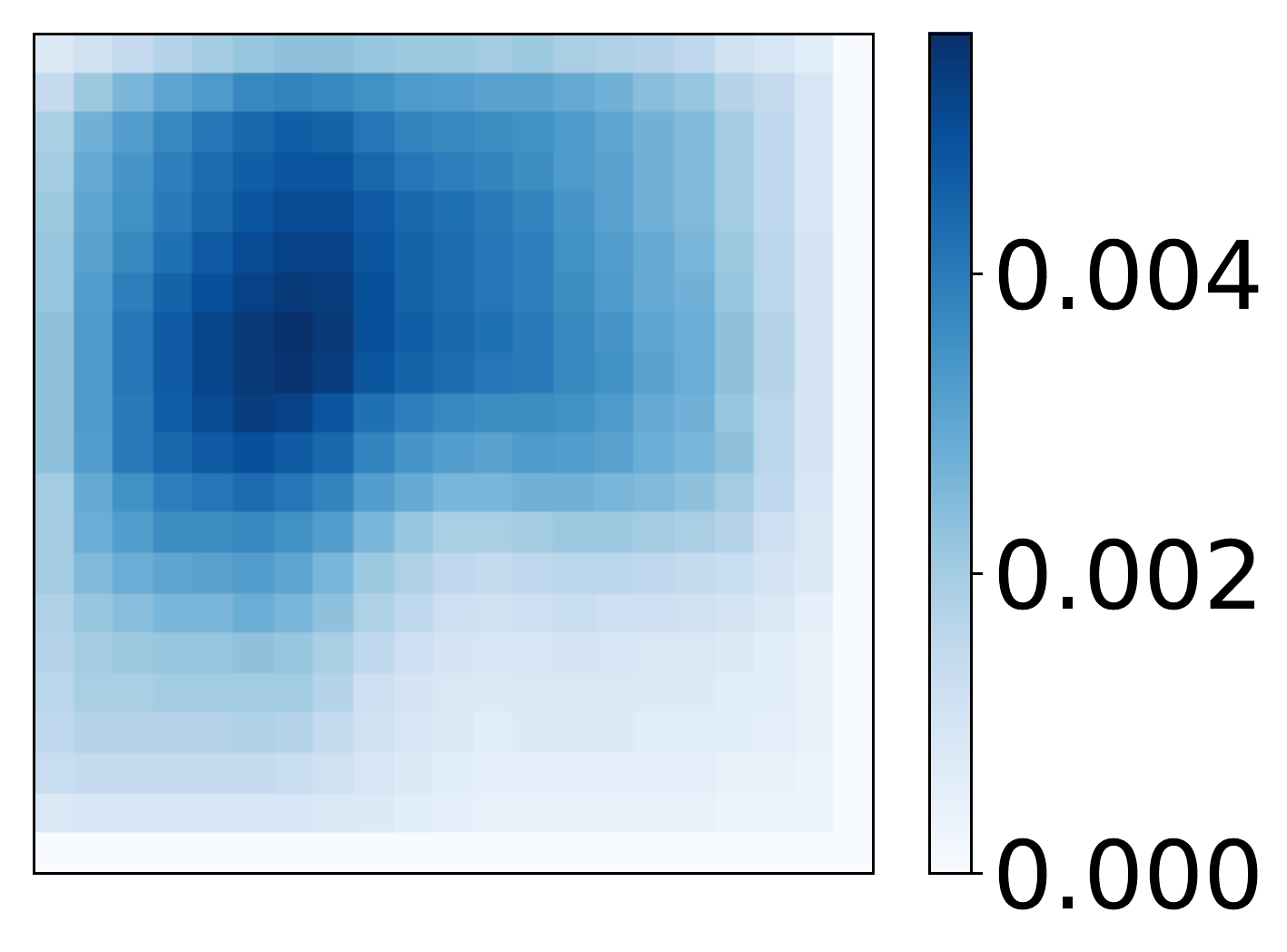}
}
\subfigure[]{
\includegraphics[width=0.17\linewidth]{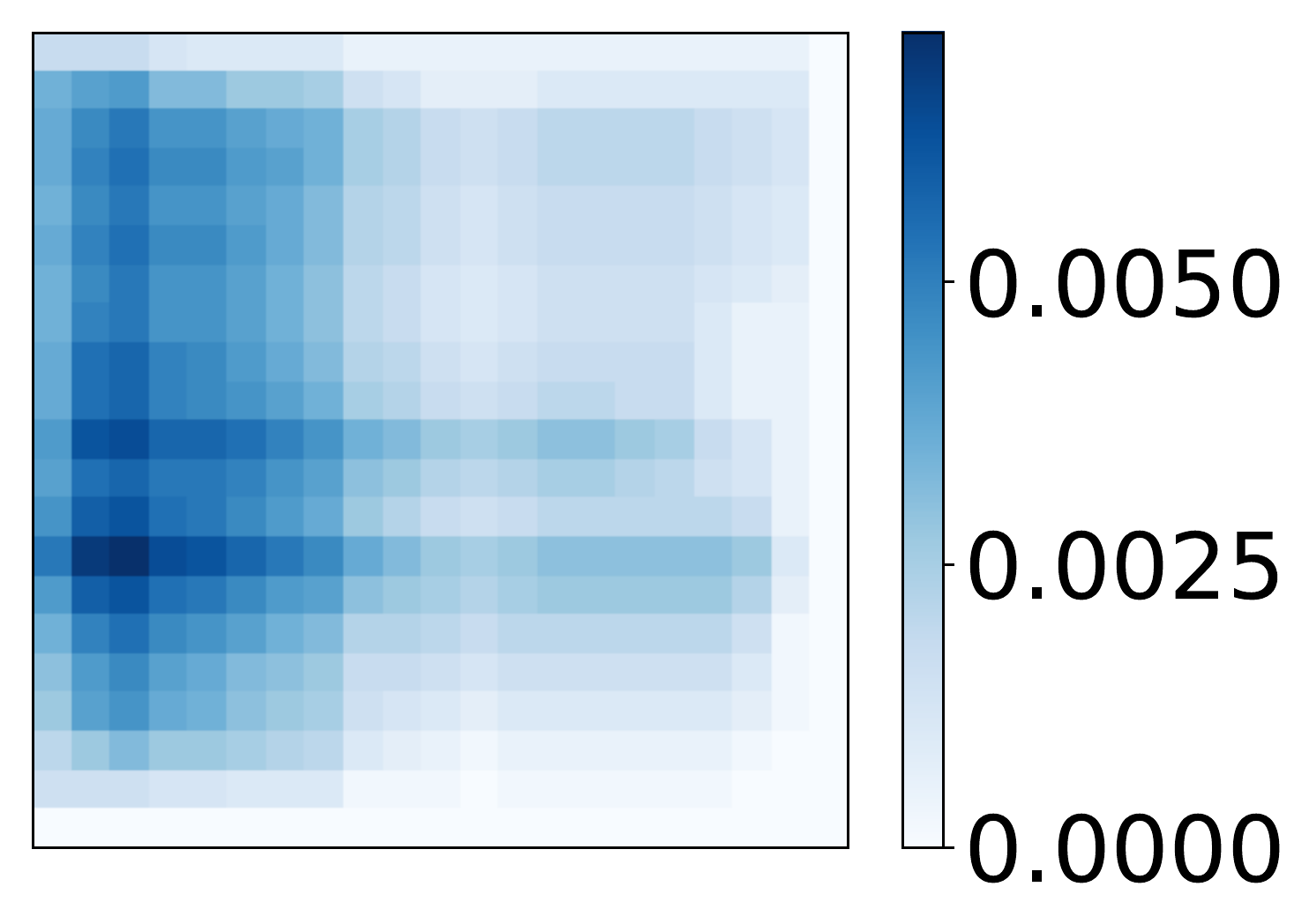}
}
\subfigure[]{
\includegraphics[width=0.17\linewidth]{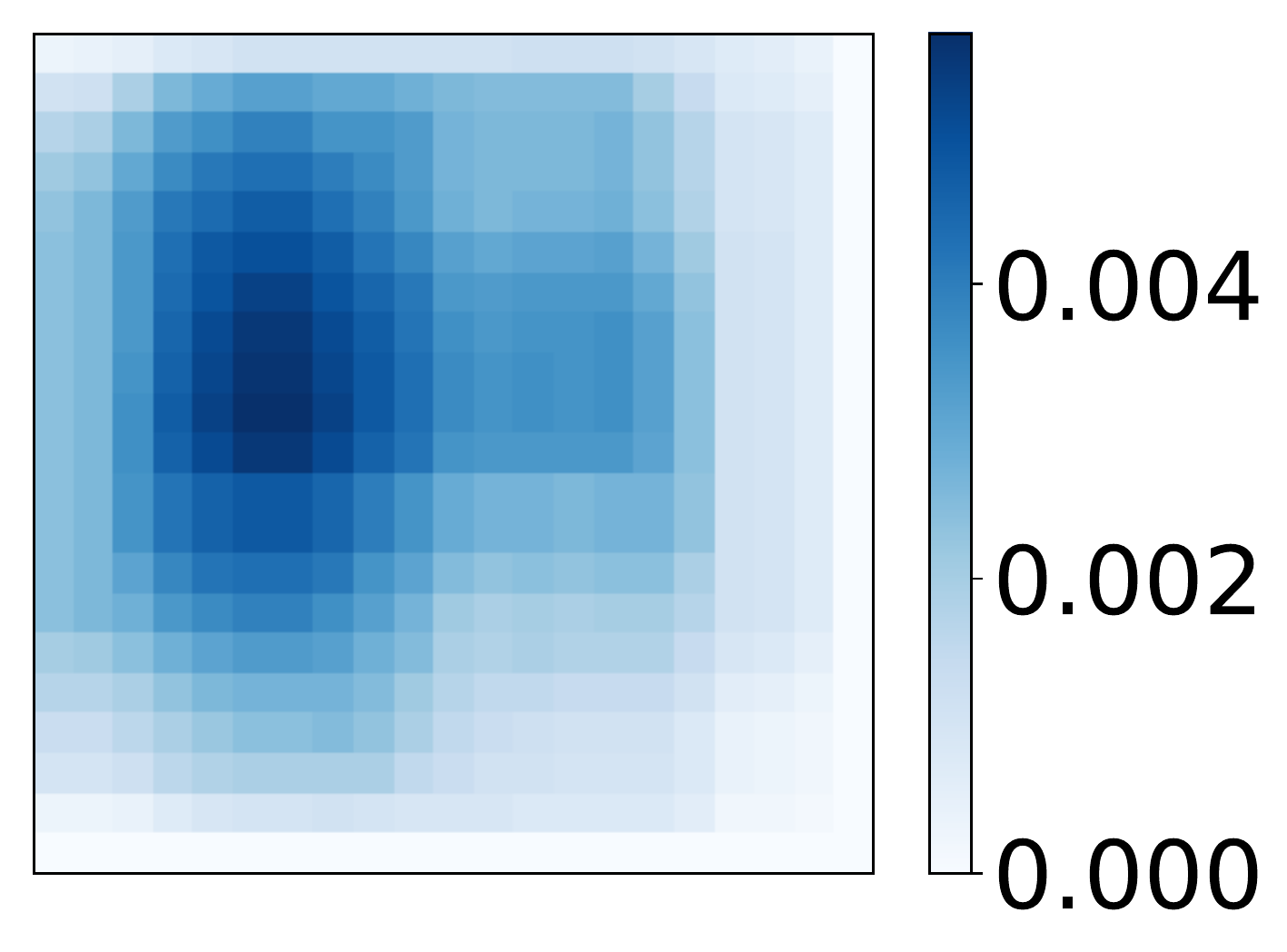}
}
\subfigure[]{
\includegraphics[width=0.17\linewidth]{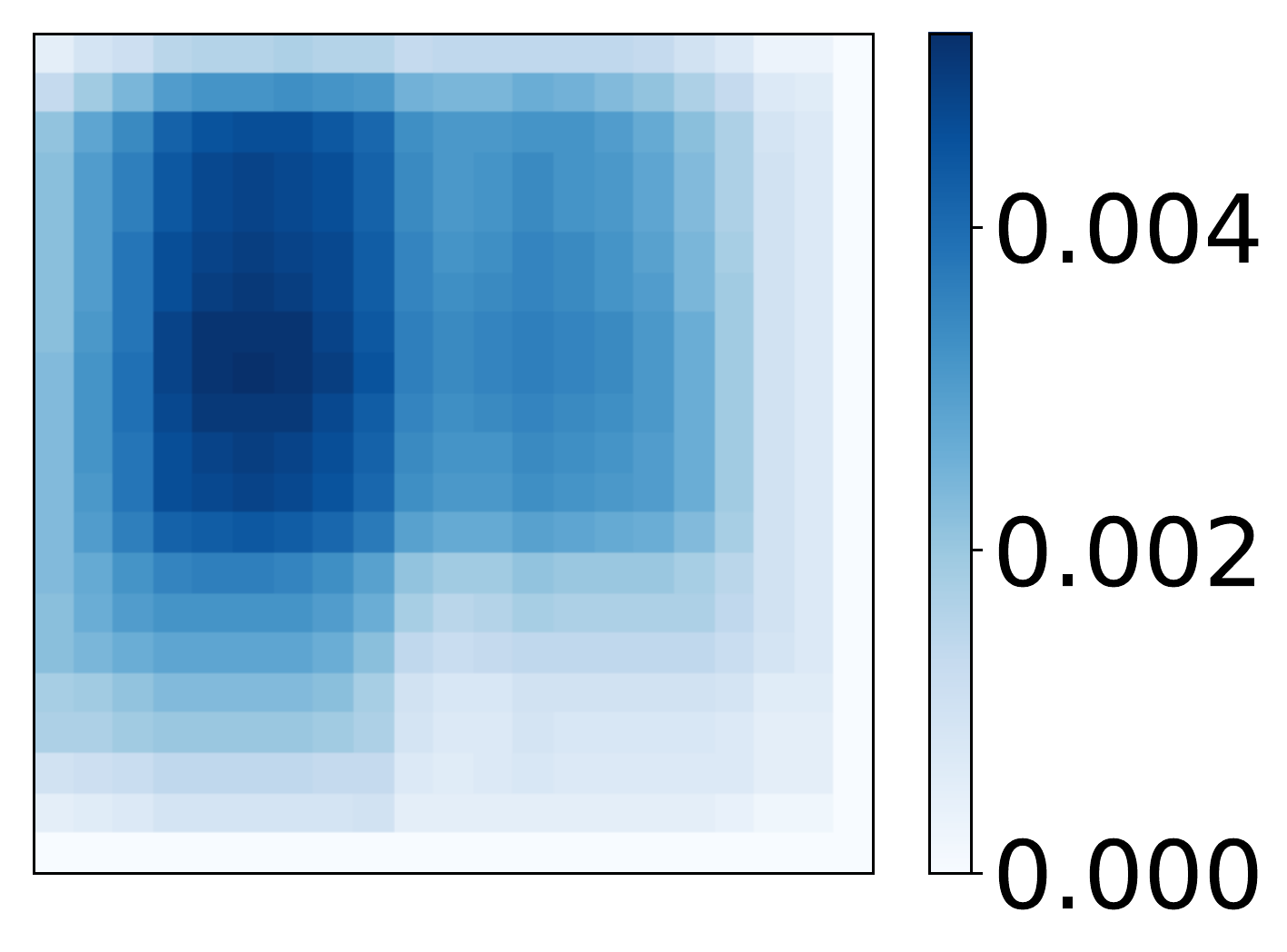}
}
\subfigure[]{
\includegraphics[width=0.17\linewidth]{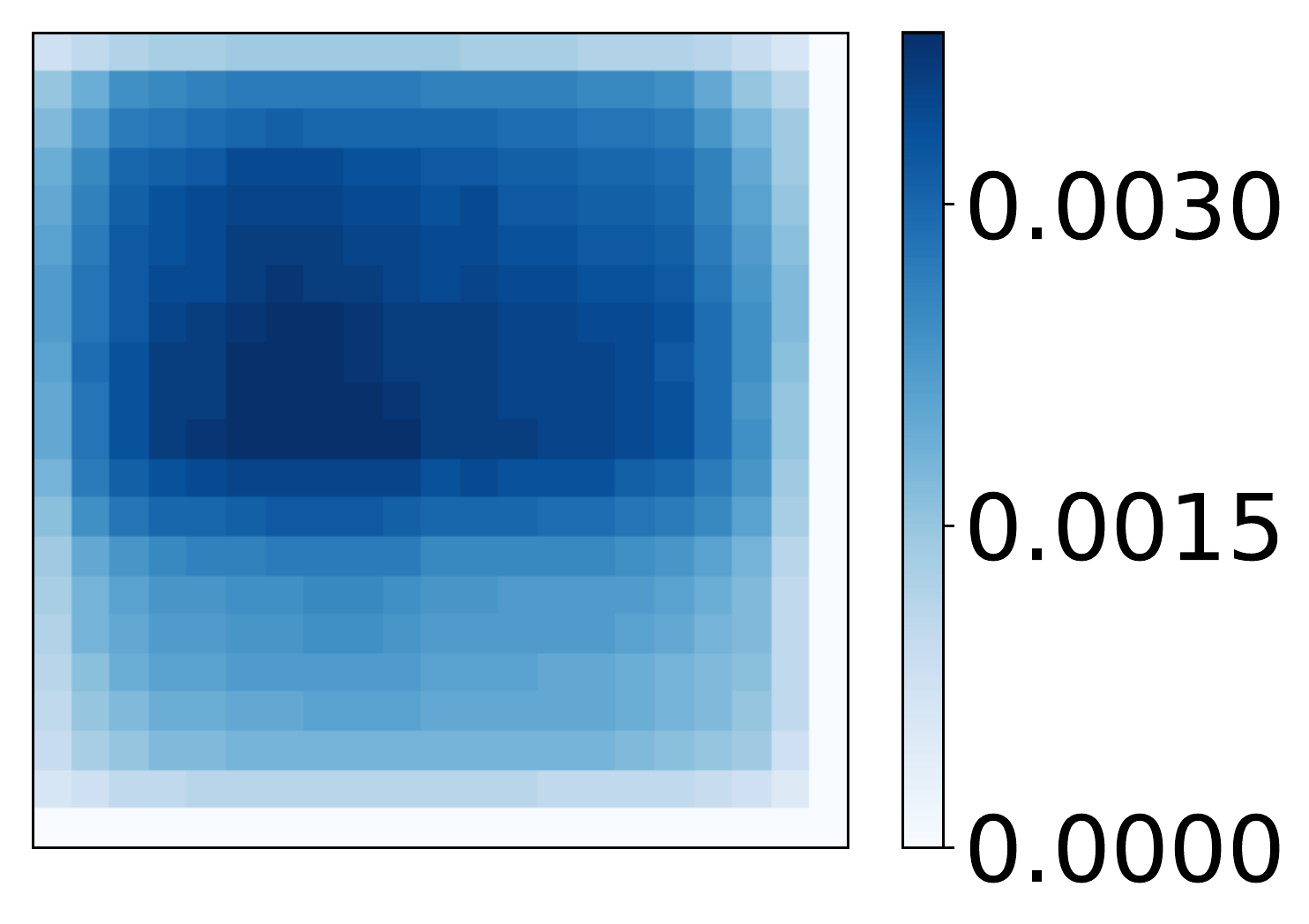}
}
\caption{(a)-(b) The distributions of element type and text length. (c)-(l) The position distributions of number, before, modifier, after, statement, single icon, bar, pie, donut, and pictograph.}
\label{fig:distribution}
\end{figure}

\paragraph{Labeling.} 
These examples are all stored as bitmap images. For each visual element, we manually label the following visual attributes: 
\begin{compactitem}
    \item Element type $t$, which can be one of the visual elements introduced in Section~\ref{subsec_visual_space}, including \emph{before}, \emph{modifier}, \emph{number}, \emph{after}, \emph{statement}, \emph{single icon}, \emph{donut}, \emph{pie}, \emph{bar} and \emph{pictograph}.
    \item Built-in attribute $b$, which is the length of characters for a textual element and the aspect ratio for a graphical element.
    \item Position $g = (x^l,y^l,x^r,y^r)$, which is the top-left coordinate and bottom-right coordinate for the element boundary.
    \item Color $c$, where we label at most two colors for an element. Specifically, we label one outline color and one fill color for each icon to ensure that its style can be matched when new infographics are generated. Besides, we label the dominant font color when there are multiple font colors in one text box.
    \item Text-specific attributes $a$, which includes the font and whether there is an italic and bold effect in the textual element\footnote{The font is recognized by using https://www.myfonts.com/WhatTheFont/.}. Specifically, we label the dominant font type and font effect (e.g., bold) when there are multiple ones in one text box.
\end{compactitem}
We treat canvas as a special graphical element and label above attributes for it. Figure~\ref{figs:dataset}(c) shows a concrete labeling result for the infographic in Figure~\ref{figs:dataset}(b). Figure~\ref{fig:distribution} shows distributions of labeling results.


Although there are other subtle visual attributes, we leave them for future work as essential visual attributes described above  can already help us generate good infographics in many cases. 
We think there are several possible ways to effectively enrich visual attributes. 
For example, labeling process is quite simple and does not require design expertise, and thus it is possible to accelerate it by crowdsourcing. 
On the other hand, with the growth of computer vision techniques, current manual labeling process can be replaced by machines~\cite{chen2019towards}, which will make the example library construction even more scalable and effective.

Copyright is another big concern of this approach, as it is a violation of intellectual property rights to fully imitate an example without owning appropriate copyrights. To resolve this issue, we leverage two datasets in our implementation. One dataset of $50$ infographics is created by our designers, which is used to retrieve examples for imitation. The other dataset of $829$ infographics is collected from the Internet and used in the non-copyright related parts of our approach, e.g., extracting the distribution of visual elements and training a recursive neural network. The copyright issue is further discussed in Section~\ref{subsec_copyright}.


\section{Retrieval}
\label{sec_retrieve}
On top of the example library, we build a retrieval system to help find appropriate examples whose designs can be potentially adapted to the input statement given by the user.

\subsection{Example Index}
\label{subsec_example_index}


Intuitively, if the input $U$ can be represented by an icon and a text statement, examples that exactly have one icon and one text box are more likely to be candidates for reuse. Moreover, if the text (or icon) of the input has a similar length (or aspect ratio) to that in an example, the design of this example will be ideal for reuse. This motivates us to index examples by \emph{visual element types} $t$ and their \emph{built-in attributes} $b$, including the number of characters for textual elements and the aspect ratio for graphical elements. Formally, for each example $D_m\in\mathbb{D}$, its index $I_m$ can be denoted as
$
    I_m = \{(t_m^n,b_m^n)\}_{n=1}^N
$, where $N$ is the number of visual elements in the example.

\subsection{Query Generation}
\begin{figure}[t]
    \centering
    \includegraphics[width=\columnwidth]{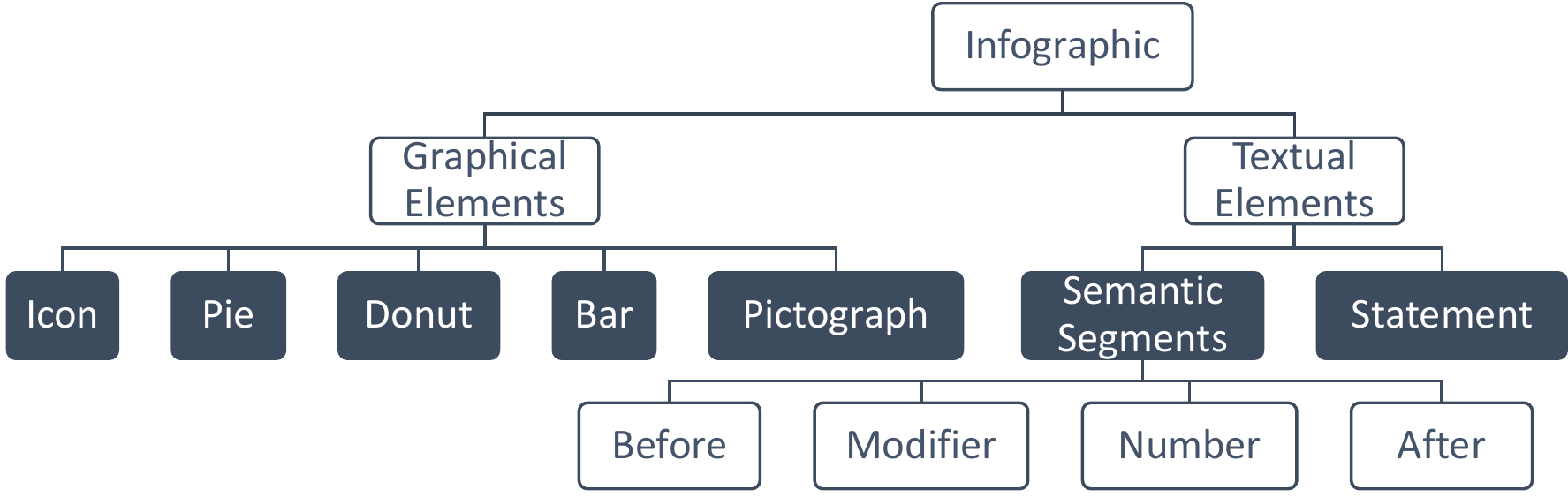}
    \caption{Illustration for design choices about visual elements.}
    \label{figs:pt}
\end{figure}

As $I_m$ is composed of visual elements types and built-in attributes, we need to transform the input $U$ to concrete queries of the same form. On the one hand, there are usually many valid ways to represent the same piece of information using infographics. For example, ``More than 74\% of users are female'' can be visualized as \emph{statement + pie} or \emph{statement+donut+icon}. On the other hand, the example library indicates the likelihood of different representations adopted in the practice. Therefore, we propose to first learn a distribution about visual elements from the example library and then generate user queries by sampling from the distribution.

\paragraph{Learning distribution about visual elements.}
Figure~\ref{figs:pt} shows design choices about visual elements. First, an infographic usually contains two main categories of visual elements, i.e., \emph{textual elements} and \emph{graphical elements}. Then, there are many design choices under these two categories (filled rectangles in Figure~\ref{figs:pt}). For \emph{textual elements}, there are two design choices, i.e., regarding the input statement as a whole and splitting the input statement into segments by semantics, denoted by \emph{statement} and \emph{semantic segments} in Figure~\ref{figs:pt}. Note that \emph{semantic segments} can be expanded to concrete visual elements, which may further yield valid combinations, e.g., \textit{number}+\textit{after} and \textit{before}+\textit{number}+\textit{after}. For \emph{graphical elements}, there are also multiple choices among the combinations over \emph{icon}, \emph{bar}, \emph{pie}, \emph{donut} and \emph{pictograph}, e.g., \textit{bar}+\textit{icon} and \textit{donut}+\textit{icon}. Formally, for a given design choice about visual elements $(u_1,\dots, u_K)$, we calculate its occurrence probability by 
$
    p(u_1, \dots, u_K) = \frac{\#(u_1,\dots,u_K)}{M}
$,
where $\#$ stands for the occurrences of $(u_1, \dots, u_K)$ in $\mathbb{D}$ of size $M$, $1\le K\le 7$, and $u_k\in\{$\emph{statement}, \emph{semantic segments}, \emph{icon}, \emph{bar}, \emph{pie}, \emph{donut}, \emph{pictograph}$\}$.

\textbf{Generating user queries.}
To ensure the diversity of generated infographics, for a given input $U$, we generate $M^\prime$ queries by following steps. First, we sample $M^\prime$ different design choices from $p$, where $p$ is the learned distribution of design choices about visual elements. Second, we specify design choices with concrete visual elements and then get their built-in attributes. Specifically, if \emph{semantic segments} exists in a query, we expand it to the corresponding visual elements, e.g., \emph{before}, \emph{modifier}, \emph{number} and \emph{after} (Section~\ref{sec_preliminary}). If \emph{icon} or \emph{pictograph} exists in a query, we leverage the result of \emph{icon generator}, which extracts representative icons according to semantics of the input statement. If \emph{pie}, \emph{donut} or \emph{bar} exists in a query, we leverage the result of \emph{chart generator}, which generates a specific chart according to the percentage in the input statement. Finally, we generate a set of queries $Q = \{Q_{m^\prime}\}_{m^\prime=1}^{M^\prime}$ where each user query corresponds to one design choice about visual elements. Formally, we encode the user query $Q_{m^\prime}$ by visual element types $\Tilde{t}_{m^\prime}^{n^\prime}$ and built-in attributes $\Tilde{b}_{m^\prime}^{n^\prime}$, i.e.,
$
    Q_{m^\prime} = \{(\Tilde{t}_{m^\prime}^{n^\prime}, \Tilde{b}_{m^\prime}^{n^\prime})\}_{n^\prime=1}^{N^\prime}
$, where $N^\prime$ is the number of visual elements in the query.

\subsection{Retrieving Strategy}
We retrieve examples by comparing the distance between example indexes and user queries. Specifically, for each query $Q_{m^\prime}\in Q$, we extract one example $\Tilde{D}\in\mathbb{D}$ whose example index has the smallest distance to the query. This will result in a set of retrieved examples with size $M^\prime$. Concretely, for a given example index $I_m$ and a user query $Q_{m^\prime}$, we first match their elements by type. Then, we calculate the distance between them by considering whether they have similar visual elements types and built-in attributes,
\begin{align}\nonumber
  S\left(I_m, Q_{m^\prime}\right) = \sum_n s_n\left(t_m^n,b_{m}^{n},\Tilde{t}_{m^\prime}^{n},\Tilde{b}_{m^\prime}^{n}\right),\;\text{where}\; 
        s_n=\begin{cases}
                    1, \;\text{if}\; t_m^n\neq \Tilde{t}_{m^\prime}^{n},\\
                    \frac{\left|b_{m}^{n}-\Tilde{b}_{m^\prime}^{n}\right|}{\Tilde{b}_{m^\prime}^{n}}, \;\text{otherwise}.
            \end{cases}
\end{align}


\section{Initialization}
\begin{figure}[t]
    \centering
    \includegraphics[width=\columnwidth]{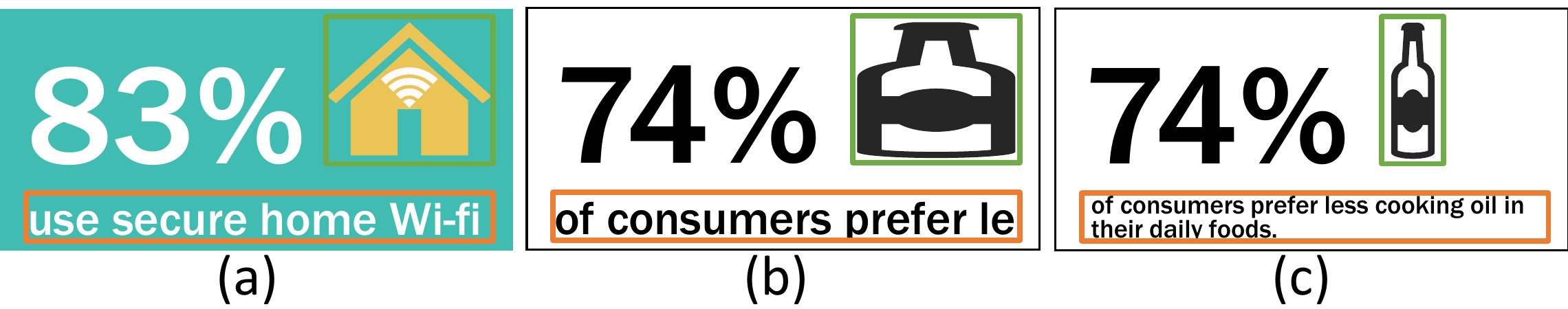}
    \caption{Example of initialization. (a) A retrieved example. (b) The initial draft generated by directly applying positions in the retrieved example. (c) The initial draft generated by scaling the graphical elements and recalculating the font size of textual elements.}
    \label{figs:bad_initialization}
\end{figure}

In the previous stage, we transform the input to a set of queries and retrieve one example for each query. In this section, for each retrieved example, we generate one initial draft by reusing its design. 

\paragraph{Position.}
For a graphical element, the aspect ratio specified in the query is usually not the same as that used in the example. Therefore, if we directly apply positions (i.e., the top-left and bottom-right coordinates) of the one in the retrieved example, the new graphical element in the initial result will be easily distorted (Figure~\ref{figs:bad_initialization}(b)). To avoid such distortion, we reuse the top-left coordinate used in the retrieved example, and recalculate the bottom-right coordinate by uniformly scaling the new graphical element to fit the space occupied by the old one (Figure~\ref{figs:bad_initialization}(c)). For a textual element, the number of characters used in the user query is usually not the same as that used in the example either. If we do not recalculate the font size, there will easily be truncated text or large white space in the text box (Figure~\ref{figs:bad_initialization}(b)). To tackle this problem, we recalculate the font size by making the font size as large as possible while maintaining the text information complete (Figure~\ref{figs:bad_initialization}(c)).

\paragraph{Color/Text-Specified Attributes.} 
For the other extracted design choices, e.g., color and font type, we directly reuse them for the visual elements specified by the corresponding query.
%


\section{Adaption}
In the previous stage, we generate an initial draft by reusing the design of the retrieved example. While the initial draft roughly conforms to design principles, spatial relationships between visual elements are not good enough in many cases. This is mainly caused by that built-in attributes in the query are often slightly different from that in the retrieved example. For instance, in Figure~\ref{figs:bad_initialization}(c), there is an excessive white space to the right side of the icon. In addition, when the text of the query is placed in the same text box of the retrieved example, the font size is too small to read. Therefore, to ensure the quality of generated infographics, it is necessary to adjust spatial relationships between visual elements in the initial draft. In this section, we introduce an adaption stage, which is a MCMC-like approach, to solve this problem. 

\subsection{MCMC-Like Approach}
\begin{figure}[t]
    \centering
    \includegraphics[width=\columnwidth]{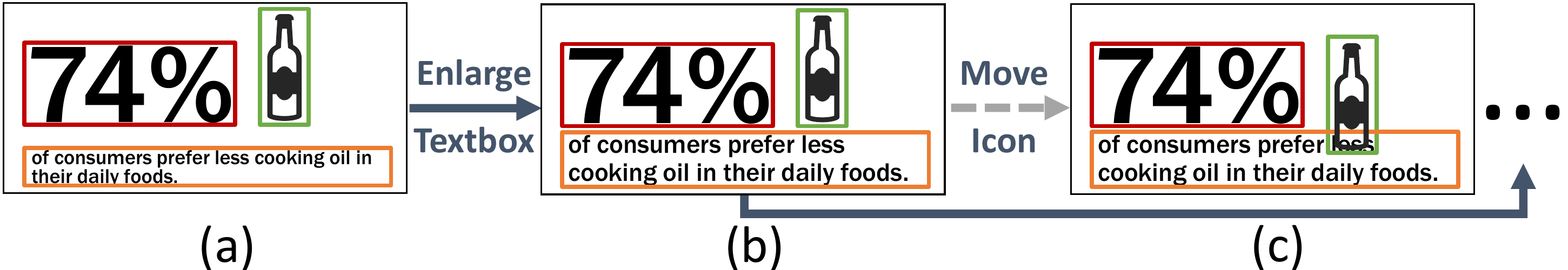}
    \caption{An illustrative example for our MCMC-like approach. (a) The initial draft. (b) A candidate design proposed by enlarging the bottom text box in (a). It is accepted as the start point for the next round of proposal. (c) A candidate design proposed by moving the icon in (b). It will be accepted with a low acceptance probability as it causes incorrect overlap. If it is not accepted, the design in (b) will continue to be used as the start point for the next round of proposal.}
    \label{figs:MCMC}
\end{figure}

As shown in Figure~\ref{figs:bad_initialization}, despite imperfectness, the initial draft provides a good start point in the complex and huge design space. Intuitively, if we search around the initial draft in the design space, there stands a good chance of finding a better design. Actually, this simple intuition is consistent with the core concept of Markov chain Monte Carlo (MCMC) methods used in traditional tasks of layout generation (e.g., graphics design or indoor scene generation~\cite{zhao2011image,qi2018human}). For these tasks, MCMC methods start from a random layout, then iteratively propose a candidate layout according to a prior proposal distribution and accept the candidate with a certain probability. After Markov chain converges, an ideal layout that obeys the target distribution is naturally achieved.

Inspired by MCMC methods, we propose a MCMC-like approach to adjust spatial relationships in the initial draft by following three steps, where the last two steps are executed iteratively until there is no better proposal. Figure~\ref{figs:MCMC} illustrates an example for our MCMC-like approach.

\paragraph{Set a Start Point.} 
As the previous stage has generated an initial draft roughly conforming to design principles, we take the initial draft as the start point instead of randomly sampling one (Figure~\ref{figs:MCMC}(a)).

\paragraph{Propose a Candidate Design.}
To adjust spatial relationships effectively, we propose candidate designs by making a small change to the current result (the initial draft for the first iteration). Spatial relationships often relate to two factors, i.e., positions and sizes of visual elements. Therefore, we design two simple dynamics corresponding to these two factors, i.e., \emph{position modification} and \emph{size modification}, and utilize them randomly to propose candidate designs. 
\begin{compactitem}
    \item \textbf{Position modification.} This dynamic randomly chooses a visual element, and proposes a new position $\Tilde{g}$ by adding a random variable $\delta g$ to the current position $g$ (i.e., $\Tilde{g} \gets g + \delta g$) of the element, where $\delta g$ follows a bivariate normal distribution.
    \item \textbf{Size modification.} This dynamic randomly chooses a visual element, samples a scale from a normal distribution, and uses the scale to resize the corresponding visual element.
\end{compactitem}

\paragraph{Compare Candidate and Current Designs.}
Once we propose a candidate design, we must determine whether to accept it for the next iteration. We adopt a similar mechanism to Metropolis-Hasting algorithm, where the candidate design is accepted with an acceptance probability $\alpha = \min\{1, d\left(D^\prime\right)/d\left(D\right)\}$.
Here $D^\prime$ and $D$ stand for the candidate and current designs respectively, and $d\left(\cdot\right)$ measures the quality of the spatial relationships in a given design. Specifically, if spatial relationships in the candidate design are better than those in the current design, i.e., $d\left(D^\prime\right)>d\left(D\right)$, we always accept the candidate design (see Figure~\ref{figs:MCMC}(b)). Otherwise, we accept the candidate design with the acceptance probability of $d\left(D^\prime\right)/d\left(D\right)$ (see Figure~\ref{figs:MCMC}(c)). The key challenge is evaluating spatial relationships of designs, i.e., estimating $d\left(\cdot\right)$.

\subsection{Recursive Neural Networks for Evaluation}\label{subsec_RNN}
The acceptance probability $\alpha$ relates to the score $d\left(\cdot\right)$, which reflects the quality of spatial relationships in a design. Typical MCMC methods use hand-crafted energy functions to evaluate spatial relationships in a layout. However, it is difficult to comprehensively and accurately represent complex spatial relationships in infographics by hand-crafted energy functions. To tackle this problem, we learn from the example library about spatial relationships evaluation. Formally, we aim to learn a mapping $h$ that takes a pair of designs $(D^\prime, D)$ as the input and predict a pair of scores $(d\left(D^\prime\right), d\left(D\right))$ as the output, i.e., $h:\left(D^\prime, D\right)\to \left(d\left(D^\prime\right), d\left(D\right)\right)$. Then, $(d\left(D^\prime\right)$ and $d\left(D\right))$ will be used to calculate $\alpha$.

To learn such a mapping, we have to address two challenges. First, we should construct a training set with effective labels from the example library. The challenge is that the example library only contains labels for visual elements and attributes while our task requires labels reflecting the quality of spatial relationships in a design. Second, we should learn a good feature representation for spatial relationships in infographics.

\subsubsection{Training Set Construction}
As examples in our library are created by professionals, we naturally hypothesize that they have good spatial relationships. Besides, we assume that, if we make random perturbations on these examples, resulting designs have worse spatial relationships than original ones. The larger the perturbation is, the worse the spatial relationships are. 

This provides us a good opportunity to construct training set effectively. Specifically, for each example $D\in \mathbb{D}$, we generate a set of new designs $\hat{D}_j$ ($j \ge 1$) through randomly choosing one visual elements and randomly changing the position or size of the labeled bounding boxes. Then, we construct training data by following two patterns. 
\begin{compactitem}
    \item Take a pair of the original design and a perturbed design (i.e., $\left(D,\hat{D}_j\right)$ or $\left(\hat{D}_j,D\right)$) as the input, and $\left(1, 0\right)$ or $\left(0, 1\right)$ as the label. Here, original designs serve as good examples. This type of training data will help our model learn to differentiate ideal and bad spatial relationships, which are more common in late iterations. 
    \item Take a pair of perturbed designs with different perturbation degrees (i.e., $\left(\hat{D}_i,\hat{D}_j\right)$ or $\left(\hat{D}_j,\hat{D}_i\right)$) as the input, and take $\left(1, 0\right)$ or $\left(0, 1\right)$ as the label. Here, the design with smaller perturbation degree is denoted as $\hat{D}_i$ and serves as the good example. This type of training data will help our model learn to pick a better design when neither of them has ideal spatial relationships, which are more common in early iterations.
\end{compactitem}

In our implementation, we generate $20000$ pairs for training and $2000$ pairs for validation by following the above two patterns.

\subsubsection{Model Architecture}
\begin{figure}[t]
    \centering
    \includegraphics[width=\columnwidth]{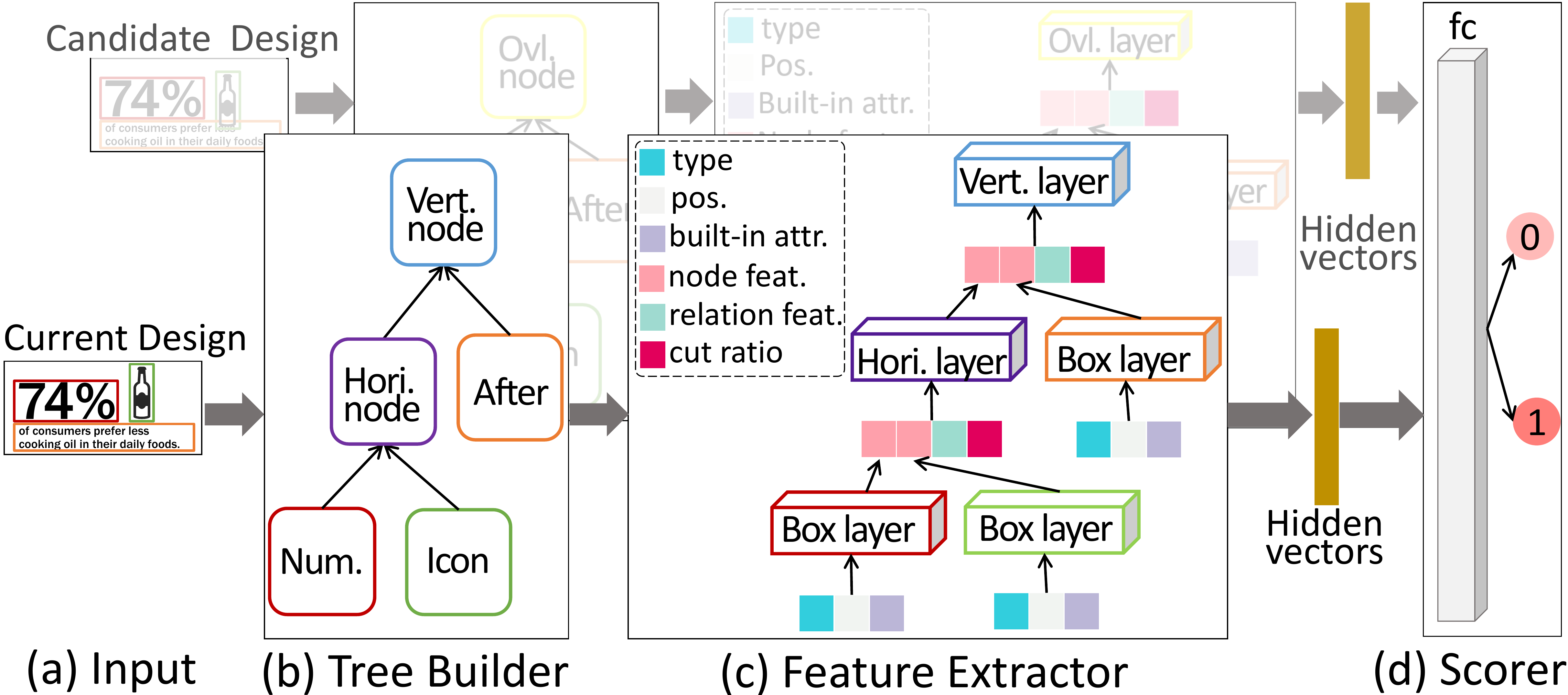}
    \caption{Concrete model architecture for spatial relationships evaluation.}
    \label{figs:layout_adaptor}
\end{figure}

The spatial relationships of infographics are inherently \emph{hierarchical}.
Taking Figure~\ref{figs:layout_adaptor}(a) as an example, its spatial structure can be represented by a tree, where the text box for \emph{number} and the \emph{single icon} are horizontally arranged and then these two elements and the text box for \emph{after} are vertically arranged.
On the other hand, recursive neural networks is a widely adopted and the state-of-the-art model to encode such hierarchical structures in existing works~\cite{li2019grains,gadi2019read}.
Therefore, we also leverage recursive neural networks, instead of convolutional neural networks~\cite{krizhevsky2012imagenet}, in our approach.


Figure~\ref{figs:layout_adaptor} illustrates the model architecture. First, the \emph{tree builder} transforms an infographic into a hierarchical tree in a top-down fashion. Then, the \emph{feature extractor}, which is a recursive neural network, encodes the infographic into hidden vectors according to the hierarchical tree in a bottom-up manner. Finally, the \emph{scorer} built upon those hidden vectors computes scores for spatial relationships in the infographic.


\paragraph{Tree Builder.}
It takes an infographic as input and outputs a hierarchical tree. Initially, the root node contains all the visual elements in the infographic. Then, we recursively split tree nodes until there is no node can be split according to following criteria.
\begin{compactitem}
    \item \textbf{Horizontal divisibility.} It means there exists a horizontal line that can divide visual elements in the current node into two groups. The current node is marked as a \emph{horizontal node} and split into two child nodes, each of which contains a group of visual elements. If there is no such cutting line, we turn to the next criterion.
    \item \textbf{Vertical divisibility.} Similar to the previous criterion, it means there exists a vertical cutting line that can divide visual elements in the current node into two groups. The current node is marked as \emph{vertical node} and is split into two child nodes. If this criterion does not apply either, we turn to the last criterion.
    \item \textbf{Indivisibility.} If there neither exists a vertical cutting line nor a horizontal cutting line, we mark this node as an \emph{indivisible node}.
\end{compactitem}

\paragraph{Feature Extractor.}
It is a recursive neural network that encodes an infographic into hidden vectors according to the hierarchical tree.

To better encode different spatial relationships, we use four distinct layers in our recursive neural network, including \emph{box layer}, \emph{horizontal layer}, \emph{vertical layer} and \emph{overlap layer}. These layers are organized according to the hierarchical tree in a bottom-up manner (Figure~\ref{figs:layout_adaptor}(c)). Each of them is a two-layer perceptron although their weights are different. Inputs for these layers are as following:

\begin{compactitem}
    \item \textbf{Box layer.} It encodes the status of the leaf node in the hierarchical tree, i.e., the visual element. The input includes: 1) the element type, a one-hot encoding, 2) the position, a vector representing the top-left and bottom-right coordinates, and 3) the built-in attribute, a scalar indicating the aspect ratio or text length.
    \item \textbf{Horizontal/vertical/overlap layers.} They encode horizontal, vertical, and overlap relationships between different nodes in the hierarchical tree, respectively. The input includes: 1) node representations for two child nodes, 2) the relation representation between two child nodes, which is a vector representing the offset of the right child relative to the left child, and 3) the cut ratio, a scalar indicating the position of the cutting line (only for the horizontal and vertical layer).
\end{compactitem}

\paragraph{Scorer.}
After the feature extractor, we get hidden vectors representing spatial relationships of the current design $D$ and the candidate design $D^\prime$, respectively. Then, we concatenate them and feed them into the scorer to get the score for spatial relationships of the infographic. Specifically, the scorer consists of a fully connected hidden layer and a Softmax layer, and is jointly trained with the aforementioned feature extractor by minimizing the cross-entropy loss.
\section{Evaluation}
\begin{figure*}[t]
    \centering
    \includegraphics[width=\textwidth]{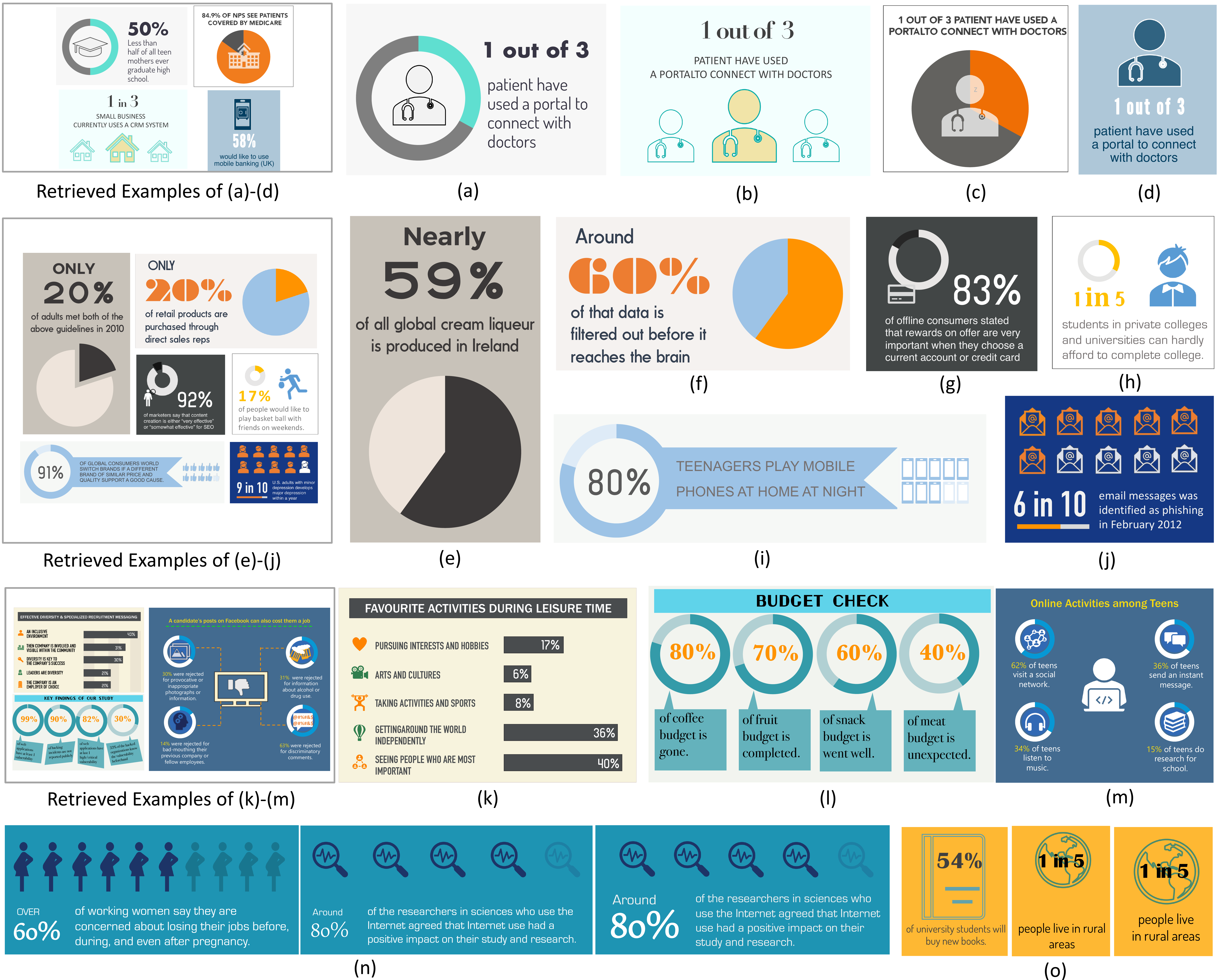}
    \caption{(a)-(j) and (k)-(m) shows infographics with single and multiple proportional facts respectively. (n)-(o) shows a triple, including the retrieved example, the initial draft generated by directly applying the retrieved example's design and the final infographic after the adaption. For better visual appearance, embellished shapes (in (i) and (l)) and the color transparency of the icon (in (c) and (g)) are added by post-editing.To avoid the copyright issue, infographics displayed here are generated by adapting the examples retrieved from a dataset created by our designers.}
    \label{figs:sample_results}
\end{figure*}


We implement a prototype system targeting widely-used proportion-related infographics.
To demonstrate the system, we use $30$ descriptions with average length of $74$. We randomly pick $5$ descriptions and generate $5$ infographics for each of them. For the remaining descriptions, we generate $1$ infographic for each of them. In this way, we generate $50$ infographics in total and show them to the experts.

Our experiments are run on a CPU Windows server (Intel Xeon E5 2.6GHz).
During the adaption, each initial draft is adjusted for $1000$ iterations, taking $3.5$s on average. Note that the adaption process is efficient because the initial draft provides a good starting point and only the inference step of the recursive neural network is used.

\subsection{Sample Infographics}

Figure~\ref{figs:sample_results} shows several typical samples from the $50$ results. Note that if there is a pictograph, we will generate multiple candidate infographics by traversing the number of icons in the pictograph from $3$ to $10$, and then leverage the model introduced in Section~\ref{subsec_RNN} to choose the one that achieves the best score. For pies, donuts, and bars, we render them by referring to the percentage provided in the input statement and reusing the annotated positions and colors of the retrieved example.

Figures~\ref{figs:sample_results}(a)-(d) are all generated from the same input utterance ``1 out of 3 patients have used a portal to connect with doctors.'' by using different queries. Specifically, we sample four different queries, including \textit{donut}+\textit{single icon}+\textit{number}+\textit{after} (Figure~\ref{figs:sample_results}(a)), \textit{pictograph}+\textit{number}+\textit{after} (Figure~\ref{figs:sample_results}(b)), \textit{pie}+\textit{single icon}+\textit{statement} (Figure~\ref{figs:sample_results}(c)) and \textit{single icon}+\textit{number}+\textit{after} (Figure~\ref{figs:sample_results}(d)). We can observe that these generated infographics have different styles in terms of the choices for visual elements, the layout, the color and the font.

Figures~\ref{figs:sample_results}(e)-(j) demonstrate that our approach provides great flexibility and diversity in dealing with different inputs. First, our approach can generate both concise and complex designs due to the natural diversity of examples. For example, Figure~\ref{figs:sample_results}(e) concisely represents information in a simple up-down structure, while Figure~\ref{figs:sample_results}(g)-(j) represent information by multiple text blocks and graphic elements in a complicated structure. Moreover, our approach is capable of highlighting the important information in various ways. For example, in Figure~\ref{figs:sample_results} (e)-(j), \emph{number} elements are set the biggest font size to attract visual attention, while in Figure~\ref{figs:sample_results}(g)-(j), icons and charts are used simultaneously to emphasize the semantic meaning and numerical information. Furthermore, our approach can create different design styles by imitating color schemes of different examples. For example, although element types are almost the same in Figure~\ref{figs:sample_results}(e) and (f), the former one shows a sense of mature and calm while the latter is more lively and vigorous. In addition, even some results have the same visual element combinations (e.g., Figure~\ref{figs:sample_results}(a), (g) and (h) all have \emph{donut}+\emph{icon}+\emph{number}+\emph{after}), they can have totally different layouts and colors if they are generated by imitating different examples.

Figures~\ref{figs:sample_results}(k)-(m) show generated compositional infographics with multiple proportional facts by imitating corresponding examples. 
To generate them, we run our approach twice. The first round generates individual proportion-related infographics by adapting an example retrieved according to element types and built-in attributes. The second round generates the compositional results by adapting an compositional example retrieved based on the number and aspect ratio of the individual infographics generated in the first round.
These compositional infographics can facilitate effective comparison among multiple proportional facts and provide a richer representation.

Figures~\ref{figs:sample_results}(n)-(o) compare the retrieved example, the initial draft generated by directly applying the retrieved example's design, and the final infographic after adaption. We observe that the initial draft roughly conforms to design principles while spatial relationships between visual elements are not good enough sometimes. For example, in Figure~\ref{figs:sample_results}(n), the font size is a little small while in Figure~\ref{figs:sample_results}(o), there is an excessive white space between the \emph{single icon} and the \emph{after}. After adaption, the spatial relationships in the initial draft are greatly improved.

\subsection{Expert Review}
To understand the effectiveness and usability of our method, we conducted an evaluation study with 4 designers. 
All of them have graduated (E1, E4) from or were enrolled (E2, E3) in professional schools of design disciplines. E1 and E4 are professional designers in a technology company. Both of them have around ten years of design experience in user interface, graphics, and video. E2 and E3 are senior graduate students and have been interns in design positions. Both of them have more than two years of design experience.

For each designer, we conduct a 60-minute interview. First, we use 10 minutes to introduce the background of this work and system workflow. Since our approach does not involve much interactions, we do not ask the designers to try out our system. Instead, we show them sample infographics generated by our approach, each containing a proportion-related natural language statement, a retrieved example, and a generated infographic. The statement and generated infographics are the input and output of our system. In particular, we show the designers the retrieved example side-by-side with the generated infographic, since we like to know their opinions about the resemblance between them. After that, a semi-structured interview was conducted to understand professional opinions on the usability and quality of our approach.

Overall, designers were very impressed by the convenience and intelligence that our approach provides to create infographics. For example, E1 mentioned that ``The capability of automatically imitating existing online examples is really smart and amazing, because it is quite time-consuming for me to do it manually.'', and E2 said that ``By intelligently imitating online examples, this approach enables so many potential ways of presenting the user information, which cannot be finished by me in a short period. It does a great job.''

Regarding how much assistance our approach can provide to users in real-world scenarios, designers thought at least two types of people would benefit from it. 
One is casual users with limited design expertise. From casual users' perspective, infographics are mainly used to enrich their reports or dashboards in informal communication. Designers believed that our generated results were good enough for them. For example, E3 mentioned that ``The generated infographics can fully express the original information and can be used directly by casual users.'' E4 said ``Compared to the uniqueness and creativity of the design, casual users usually prefer to follow popular designs, which is exactly what this approach provides.''
The other one is professional designers themselves. While our approach initially targets casual users, all the designers express the appreciation for the help our approach could provide for them as well. 
As admitted by all the designers, they had to immerse themselves with myriad examples to get inspirations during the design process. They usually login to multiple design websites and manually check which design satisfies their requirements.
First, experts really appreciated the capability of automatically retrieving the examples which meet user requirements by our approach. ``When I propose a design to a client, if no existing ones are referenced and attached, the proposal will be rejected with high probability. I have to manually search the Internet to find references. I think the retrieval stage in this approach can save my time on finding intent references'', said one of the designers. E3 said, ``Sometimes I even do not have clear thought about my designs. This approach can provide me various previews about different designs on my data. It is really informative.'' 
Moreover, designers thought the adaption stage by itself was very helpful. For example, E2 commented, ``When I want to reuse some designs, I have to manually adjust visual elements to make it suitable for my data. It is amazing that the adaption stage can automatically do it. I think it would save much time if I could make adjustment on the automatically-adapted version provided by this approach.''. 
In addition, designers further agree that the generated results could be used as their initial design drafts. E4 said, ``Even from a professional point of view, it can definitely be used as an initial design draft for designers, which can save us lots of time and efforts.''.

In terms of the quality of our generated infographics, four experts rate the results separately and think that roughly $68\%$ of them are reasonable. 
The comments of designers mainly center on three aspects. First, for most exhibited cases, the designers feel that they were flexible and diverse enough. Two designers commented regarding the element combination that ``This approach provides the flexibility of composing various visual elements to create a complex infographic.'' (E4), and ``overall it is very diverse, and each case shows a unique combination of elements.'' (E2). Another designer also commented on the layout of the generated infographics: ``even though the element combinations are similar, it generates quite different infographics by laying elements out in different ways.'' (E1). Second, almost all designers had said that the important characteristics of the example were completely preserved by our approach. For example, E1 said that ``the generated infographic looks very similar to that of the retrieved example. Only by scrutiny, could I tell the minor differences''. Finally, all the designers thought that generated results accurately convey the original information. E2 commented that ``The key information of user input, especially the numerical information, is correctly highlighted by using a larger font size or leveraging data driven graphics like bars, pies and donuts.''.

We also receive suggestions that implied further necessary improvements to our approach. 
First, it is suggested to use icon combination instead of a single icon to represent the semantic meaning of the input statement. For example, when seeing a key phrase ``phishing mail'' in the input statement, E1 expressed the intent to compound a ``fish'' icon and an ``email'' icon together to convey such information. We think the latest work ICONATE~\cite{zhao2020iconate} can be integrated to our approach to improve performance.
Moreover, designers also made suggestions on further adding a reference line in bar charts used in our results. ``Although this is a trivial detail, the reference line makes multiple bar charts easier to be understood.'' said by one of designers.





\section{Discussion}
\label{sec_discussion}
\subsection{Example-Based Infographic Generation}

Creating infographics is non-trivial, which requires tremendous design expertise and is time-consuming. Therefore, automating the generation process of infographics is very meaningful and useful.
Previous studies try to automatically generate infographics by using predefined blueprints. However, the deficiency of blueprints largely constrains the diversity of generated results. 
By contrast, there are enormous infographics on the Internet.
This motivates us to consider example-based approach, i.e., generating infographics by imitating existing infographics. 
One concern for this approach is that crowdsourced examples might not always fit right design principles and thus may have negative influence on the generation performance. To alleviate such influence, in our implementation, we control the quality of crowdsourced examples by letting annotators discard the designs with obvious errors during the labeling process and collecting examples from professional websites.

Moreover, in real scenarios, both designers and casual users get inspired from existing infographics. For designers, they usually explore lots of online examples to learn advanced design principles and accumulate design materials, which will be used in their future designs. For casual users, who have no design background, they even have strong desires to directly reuse a favorite existing infographic on their own data. These scenarios make us believe that our example-based approach is useful for casual users and even potentially helpful for designers.

In addition, we aim to use the automatic generation of infographics as the assistance but not the replacement for humans. On the one hand, it is almost impossible to automatically generate a perfect infographic. On the other hand, the design process is subjective. Therefore, human interaction is indispensable during the design process. Our approach can generate editable infographics that roughly conform to design principles and meet user requirements, based on which people can easily get desirable infographics by polishing and fine-tuning.

\subsection{Copyright Issue}\label{subsec_copyright}
Copyright is a big concern for our example-based approach.
It is a violation of intellectual property rights to fully imitate an example without owning appropriate copyrights. Practically, although designers explore online examples for inspirations, they hardly adopt every design aspect from a single example.
During the interview, the designers mentioned that may largely reuse high-level designs, e.g., layouts and colors, but they will rethink details, e.g., shading and textures, based on their personal tastes.
In this work, to demonstrate the capability of our approach, the algorithm is designed to imitate examples as much as possible.
There are several solutions to deploy our technique in production.
The simplest way is to directly use an example library with proper copyrights.
As infographics are often easier to purchase than blueprints, it may still achieve a good diversity.
In addition, as our approach contains two consecutive stages, it is also possible to allow users to directly provide an example with a correct copyright for the \textit{adaption} stage.
Another promising solution is to combine designs from multiple examples, instead of reusing every aspect from a single one.
We believe this solution is more aligned with the common practice of designers, and plan to improve our approach in this respect.


\subsection{Opportunities for New Usage}
We think there are at least three opportunities for new usage.

\paragraph{From Proportion Infographics to Others.}
Although this work focuses on proportion-related infographics, it is possible to generalize our approach to more types of infographics, e.g., time-line, location, or process. In our approach, both the retrieval and adaption stage are proposed independent of the infographics type. Thus, it can accommodate other types of infographics once the visual elements are redefined and an annotated example library are provided.

\paragraph{Customized Searching Tool for Infographics.}
The \textit{retrieval} stage in our approach can be extended to a customized searching tool for infographics. In practice, people often intend to find infographics with certain kinds of visual attributes. However, they usually cannot get satisfactory results by typing descriptions in traditional search engines, e.g., Google or Bing. For example, when people want to get inspiration from existing infographics about how to design infographics with unusual canvas size (like $5:1$), traditional search engines can hardly provide related infographics. Thus, a customized searching tool for infographics, which indexes visual attributes of infographics, is of great value. The retrieval stage of this work has already indexed some essential visual attributes. A extended version of it can greatly improve the searching efficiency for infographics.

\paragraph{Automatic Layout Adaption Tool.}
Techniques used in our \textit{adaption} stage can be potentially integrated into mainstream software to offer automatic refinement. To lower the authoring barrier, commercial software (e.g., PowerPoint or Adobe Illustrator) usually provide predefined blueprints to users. However, the user data usually cannot be perfectly fitted into these predefined blueprints, requiring tedious refinement after applying blueprints. If users can make edits based on designs that have been automatically adjusted by our adaption techniques, both of authoring efficiency and user experience can be improved.

\subsection{Limitations}

There are several limitations of our approach. 
First, the implementation for our approach is limited to proportion-related infographics although it can be potentially generalized to other types of infographics.
Second, when constructing the example library, visual elements and attributes can be further enriched to improve the performance of retrieval. For instance, various properties of charts can be added, e.g., the inner radius of donut charts or explode slices in pie charts. For another instance, embellished shapes can be added, e.g., stars, banners or speech/thought bubbles.
Third, the current retrieval stage only considers attributes of candidate visual elements but ignores the user preference on visual styles (e.g., the concise style, the lively colors, or the landscape layout). In the future, we can index examples by their visual styles and improve the retrieval strategy to consider the consistence between the user preference and the example's visual style.
Fourth, the adaption efficiency can be potentially improved by designing more efficient strategies to propose candidate designs.
In addition, we only adjust spatial relationships as it is the most easily disturbed attributes after the initialization. Other attributes can also be adjusted to further boost the performance.

\section{Conclusion and Future Work}
In this paper, we introduce retrieve-then-adapt, an automatic approach to generate infographics by imitating online examples. 
Specifically, we present a retrieval stage that indexes and retrieves online examples based on the element types and build-in attributes, and an adaption stage that automatically adjust spatial relationships by a MCMC-like approach to make the retrieved example's design more suitable for the user information. We demonstrate the expressiveness and usability of our approach through sample results and expert reviews. We believe this work opens up a new paradigm for automatically generating infographics, i.e., example-based approach. We plan to further extend this framework by considering other types of infographics, considering more design choices and mixing designs of different examples, in order to better meet users' growing demands.


\bibliographystyle{abbrv}
\bibliography{main}

\clearpage
\appendix
\section{Failure Cases}
\begin{figure}[t]
    \centering
    \includegraphics[width=\textwidth]{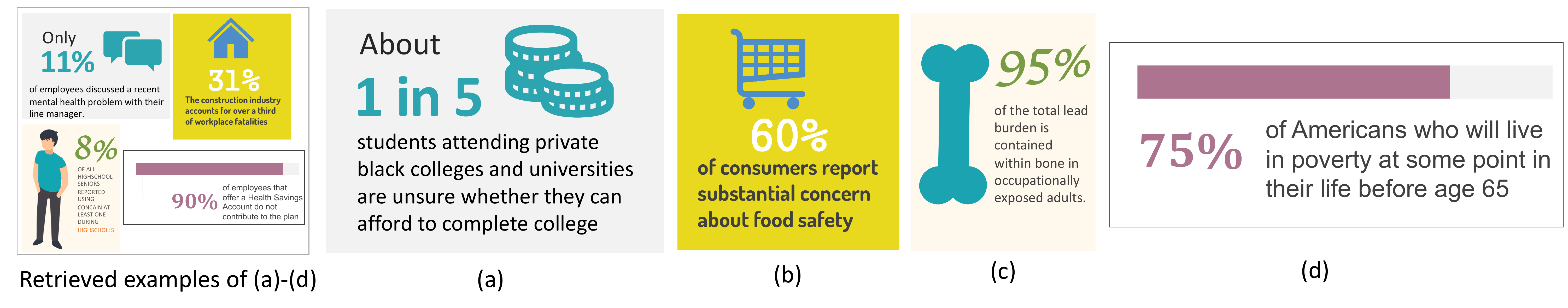}
    \caption{Failure cases.}
    \label{fig:failures}
\end{figure}

We also observe some failure cases during the evaluation. Figure~\ref{fig:failures} shows examples for common failures.

One typical failure is that the generated infographic does not meet people's aesthetic. For example, Figure~\ref{fig:failures}(a) shows a `green' coin because the color is directly inherited from the retrieved example. However, the gold or the silver is more appropriate considering the semantic meaning of the coins. Since several techniques~\cite{bahng2018coloring,lin2013selecting} have been investigated to identify semantically relevant colors from natural language statements, we expect to improve the colorization by integrating such techniques into our approach. For another example, in Figure~\ref{fig:failures}(b), there is a slight misalignment between the icon and the texts, which makes the generated infographic look less neat. Such misalignment is caused by that the evaluation model makes wrong predictions during the adaption. 
We think there are at least two possible ways to alleviate the misalignment problem. On the one hand, the performance of the evaluation model can be improved by feeding more training data.
One the other hand, a large proportion of these slight misalignments can be easily solved by post-processing generated infographics, i.e., forcing two elements to be aligned when the gap between their edges is minute.

Besides, another main failure is that the visual appearance of the generated infographic is not identical to that of the retrieved example. For the retrieved example of Figure~\ref{fig:failures}(c), its icon has rich color and it looks appealing. However, as we only inherits the dominant color of the icon, the generated infographics in Figure~\ref{fig:failures}(c) looks much plain. We think labeling as much colors as possible will not fundamentally solve this problem. One possible solution is to leverage computer vision techniques to automatically colorize the icon~\cite{sun2019adversarial}. For another instance, in the retrieved example of Figure \ref{fig:failures}(d), there is a connecting line between the bar chart and the number `90\%'. In the generated infographic, there is no such line as we do not label it during the example library construction. 
We can observe that although the generated result is a reasonable infographics after adapting the spatial relationships, its overall visual appearance is different with that of the retrieved example. We believe that when more design choices are added into our system, such visual appearance gap can be gradually narrowed down.

%

\end{document}